\documentclass{llncs}
\usepackage{amsmath} 
\usepackage{amssymb,graphicx,float}
\usepackage{algorithm}
\usepackage{algpseudocode}
\usepackage{algorithmicx}
\usepackage{geometry} 
\newtheorem{prop}{Proposition}
\newtheorem{obs}{Observation}
\title {An Efficient Algorithm for the Proximity Connected Two Center Problem}
\author{Binay Bhattacharya, Amirhossein Mozafari\thanks{Corresponding author}\and
Thomas C. Shermer }
\institute{School of Computing Science\\ Simon Fraser University\\\email{\{binay,amozafar,shermer\}@sfu.ca}}
\begin{document}
\maketitle
\begin{abstract}
Given a set $P$ of $n$ points in the plane, the $k$-center problem is to find $k$ congruent disks of minimum possible radius such that their union covers all the points in $P$. The $2$-center problem is a special case of the $k$-center problem that has been extensively studied in the recent past \cite{CAHN,SH,HT}. In this paper, we consider a generalized version of the $2$-center problem called \textit{proximity connected} $2$-center (PCTC) problem. In this problem, we are also given a parameter $\delta\geq 0$ and we have the additional constraint that the distance between the centers of the disks should be at most $\delta$. Note that when $\delta=0$, the PCTC problem is reduced to the $1$-center(minimum enclosing disk) problem and when $\delta$ tends to infinity, it is reduced to the $2$-center problem. The PCTC problem first appeared in the context of wireless networks in 1992 \cite{ACN0}, but obtaining a nontrivial deterministic algorithm for the problem remained open. In this paper, we resolve this open problem by providing a deterministic $O(n^2\log n)$ time algorithm for the problem.
\end{abstract}
\section{Introduction}
The $k$-center problem in the plane is a fundamental facility-location problem in which we are given a set of $n$ demand points $P$ and we are going to find a set $S$ of $k$ center points such that $cost(S):=\max_{p\in P}\min_{s\in S}dist(p,s)$ is minimized ($dist(p,s)$ is the Euclidean distance between $p$ and $s$). The $k$-center problem is known to be NP-hard~\cite{ASH2}. However, there is a simple greedy $2$-approximation algorithm for the problem which can not be improved unless $P=NP$~\cite{ASH2}. So, the studies on the problem went in the direction of obtaining polynomial-time algorithms where $k$ is not considered as a part of the problem input. As an example, in 2002,
Agarwal and Procopiuc~\cite{AP} gave a $n^{O(\sqrt{k})}$ time algorithm to solve the $k$-center problem. Solving the problem for specific values of $k$ like $k=1$ and $k=2$ received attention due to the geometric properties that can be applied to solve these problems efficiently. The $1$-center problem is indeed equivalent to the problem of covering $P$ with a disk with minimum area. This problem is also called the \textit{minimum enclosing disk (MED)} problem. In 1983, Megiddo~\cite{MG} used the prune and search technique to give an optimal linear time algorithm to solve the MED problem.\par
For $k=2$, Drenzer~\cite{DRENZ} gave the first nontrivial algorithm for
the problem with $O(n^3)$ time complexity. Later in 1994, Agarwal and Sharir~\cite{ASH} improved the time complexity for the problem to $O(n^2\log^3 n)$. In 1996, Eppstein~\cite{EPS} gave a randomized algorithm for the problem with $O(n\log^2 n)$ expected running time. In 1997, Katz and Sharir~\cite{KSH} proposed the novel expander-based parametric search technique and showed that applying it to the $2$-center problem using the $O(n^2)$ time feasibility test of Hershberger~\cite{HFT}, gives an $O(n^2\log^3 n)$ time algorithm for the problem. Later in the year, Sharir~\cite{SH} designed an $O(n\log^3 n)$ time algorithm for the decision version of the $2$-center problem using the breakthrough observation of breaking the problem into three separate cases(far distant, distant and nearby cases). Next, he parallelized the decision algorithm and put it into the Megiddo's parametric search schema~\cite{MGPS} to obtain an $O(n\log^9 n)$ time algorithm. Soon, it turned out that solving the problem in the nearby case is the bottleneck to reduce the time complexity. Later, Sharir's running time was improved by Chan~\cite{CHN} and Wang~\cite{HT} to $O(n\log^2n\log^2\log n)$ and $O(n\log^2 n)$ respectively. Very recently, Choi and Ahn~\cite{CAHN} (independently Cho and Oh~\cite{CHOO}) obtained an $O(n\log n)$ time algorithm for the nearby case which led to an optimal $O(n\log n)$ time algorithm for the $2$-center problem. \par
We say that a set $S$ of $k$ center points in the plane satisfies the \textit{proximity connectedness condition (PCC)} with respect to a parameter $\delta$ if the $\delta$-distance graph of $S$ (the graph with vertex set $S$ such that there is an edge between two vertices if and only if the distance between them is at most $\delta$) is connected. The \textit{proximity connected $k$-center} problem is defined as a generalized version of the $k$-center problem for which, in addition to $P$, a parameter $\delta\geq 0$ is also given. The objective is to find $k$ center points $S$ such that $S$ satisfies the PCC and $cost(S)\leq cost(S')$ for any $k$ center points $S'$ that satisfies PCC ($cost(S)$ is the same cost as in the $k$-center problem). Note that when $\delta$ tends to zero (resp. infinity), the problem reduces to the $1$-center (resp. $k$-center) problem. Also, when $\delta$ tends to zero and $k$ tends to infinity the problem becomes the Euclidean Steiner tree problem (connecting the points of $P$ by lines of minimum total length in such a way that any two points can be connected by the lines). This is because in this configuration, the centers should be placed along the lines of the minimum Steiner tree in order to minimize the cost. The Euclidean Steiner tree problem is also NP-hard but it has a PTAS approximation algorithm~\cite{AROR}. \par
In practice, the parameter $\delta$ usually specifies the range for which one center can communicate with other centers. So, when $S$ satisfies the PCC, any pair of centers can communicate with each other via the other centers. The proximity connected $2$-center (PCTC) problem first emerged in the works of Huang~\cite{ACN0} in 1992 while he was studying packet radio networks. In the network terminology, the PCTC problem is the problem of locating two wireless devices as close as possible to the demand points $P$ such that they can send/receive messages between each other. He originally gave an $O(n^5)$ time algorithm for the $2$-center problem having proximity constraints between their centers. Later in 2003, Huang \textit{et al.}~\cite{ACN1} studied a very close problem to the PCTC problem called \textit{$\alpha$-connected} $2$-center problem. In this problem, instead of $\delta$, a parameter $0\leq\alpha\leq 1$ is given and the distance between the center of the disks should be at most $2(1-\alpha)r$ where $r$ is the radius of the disks. They gave an $O(n^2\log^2n)$ time algorithm for the decision version (given an $r$ whether it is possible to cover the points with two disks of radius $r$ satisfying the desired conditions) of the problem. Note that this problem is a special case of the PCTC decision problem where $\delta= 2(1-\alpha)r$. Later in 2006, they gave a randomized algorithm with the same $O(n^2\log^2n)$ expected running time to solve the \textit{$\alpha$-connected} $2$-center problem~\cite{ACN2}. In this paper, we consider the PCTC problem and propose a deterministic $O(n^2\log n)$ time algorithm for it.\par
Here, we need to mention that although we use Sharir's observation~\cite{SH} of breaking the problem into three different cases(far distant, distant and nearby), the reason we can't get a sub-quadratic algorithm like~\cite{SH,CHN,HT,CAHN} is that the PCTC problem is structurally different from the $2$-center problem. In the $2$-center problem, the optimal cost is determined by at most three points of $P$~\cite{SH} while in the PCTC problem the cost may be determined by more than three points because of the PCC. This means that our search space has a higher dimension than the search space of the $2$-center problem. Also, all the sub-quadratic algorithms for the $2$-center problem use Megiddo's~\cite{MGPS} or Cole's~\cite{CL} parametric search schema to reduce the time complexity which makes the resulting algorithm impractical~\cite{ASH} while our algorithm exploits the geometric properties of the problem which make it straightforward to be implemented using standard data structures in computational geometry. \par
A \textit{solution} for a given PCTC problem instance is defined as a pair of disks whose centers satisfy the PCC and their union covers $P$. We call a disk with the larger (or equal) radius the \textit{determining disk} of the solution and its radius the \textit{cost} of the solution. An \textit{optimal solution} is a solution with minimum cost among the set of all solutions for the problem. Note that there might be an infinite number of optimal solutions with different pairs of radii because we have freedom on the smaller disk. So, we try to find an optimal solution such that the radius of its smaller disk is minimum among all optimal solutions. We call such a solution a \textit{best optimal solution (BOS)} for the problem. Therefore, if the problem has more than one  BOS, they would have the same pair of radii. We can also compare two solutions $S_1$ and $S_2$ as follows: we say that $S_1$ is a better solution than $S_2$ if $cost(S_1)<cost(S_2)$ and if $cost(S_1)=cost(S_2)$, the radius of the non-determining disk of $S_1$ is smaller than the radius of the non-determining disk of $S_2$. In this paper, our algorithm not only gives us an optimal solution but it computes a BOS for the problem. 
\section {Preliminaries and Definitions}
Let $(P,\delta)$ be the given PCTC problem instance where $P$ is a set of $n$ demand points in the plane and $\delta$ is a given non-negative number. We assume that the points are in general position, by which we mean no four points of $P$ lie on a circle. Let $(P_1,P_2)$ be a partition of $P$ obtained by dividing the plane by a line or two half-lines from a point (henceforth, when we use the term \textit{partition of the plane}, we mean a partition that satisfies this condition). We say that a pair of disks $(D_1,D_2)$ with centers $(c_1,c_2)$ respectively is a \textit{solution for the partition} if $D_1$ covers $P_1$, $D_2$ covers $P_2$ and $dist(c_1,c_2)\leq \delta$. \textit{Optimal} and \textit{best optimal solutions (BOSs)} for the partition are defined similarly. Let $(D^*_1,D^*_2)$ be a BOS for the partition with centers $(c^*_1,c^*_2)$ respectively. We say that a point $p\in P_1$ is a \textit{dominating point} of $D^*_1$ if $(D^*_1,D^*_2)$ is not a BOS for the partition $(P_1\setminus p, P_2)$. The dominating points of $D^*_2$ are defined similarly. Note that the dominating points of $D^*_1$ and $D^*_2$ are on their boundaries. By assuming that the points are in general position, if $D^*_1$(resp. $D^*_2$) is the MED of $P_1$(resp. $P_2$), its dominating points are either three points on the boundary such that their induced triangle contains $c^*_1$(resp. $c^*_2$) or two points on the boundary such that their connecting segment passes through $c^*_1$(resp. $c^*_2$). In order to simplify the presentation of our algorithm, in the latter case, we consider one of the dominating points as two infinitely close points and so, if $D^*_1$ or $D^*_2$ is the MED of their corresponding points, we assume that it has exactly three dominating points. Similarly, if $D^*_1$(resp. $D^*_2$) is not the MED of $P_1$(resp. $P_2$), in the case that it only has one dominating point, we can consider it as two infinitely close points. But, if it has three points on its boundary such that their induced triangle does not contain $c^*_1$, we might have no dominating point for $D^*_1$. We can assume that such a situation never happens by slightly perturbing the points. So, henceforth, if $D^*_1$(resp. $D^*_2$) is not a MED, we assume that it has exactly two dominating points.\par
We call the problem of computing a BOS for a given partition $(P_1,P_2)$ the \textit{restricted PCTC problem}. In the next section, we show that how we can solve the restricted PCTC using the \textit{intersection hulls} and the \textit{farthest-point Voronoi diagram}s of $P_1$ and $P_2$ (the intersection hull of a set of points with respect to some radius $r$ is defined as the intersection of all disks of radius $r$ around the points of the set). See Appendix A for a review on {farthest-point Voronoi diagram}s and intersection hulls and their properties.
\section{Computing a BOS for a Partition}
Let $(P_1,P_2)$ be a given partition. First, we compute the minimum enclosing disks $D_1^{**}$ and $D_2^{**}$ for $P_1$ and $P_2$ respectively. This can be done in linear time due to Megiddo's algorithm \cite{MG}. Also let $c_1^{**}$ and $c_2^{**}$ be the centers of $D_1^{**}$ and $D_2^{**}$ respectively. Now, if $dist(c_1^{**},c_2^{**})$(the distance between $c_1^{**}$ and $c_2^{**}$) is at most $D$, then we are done and $(D_1^{**},D_2^{**})$ is a BOS for the partition. Otherwise, we have the following proposition:
\begin{prop}
If $dist(c_1^{**},c_2^{**})>\delta$ then for any BOS $(D_1^*,D_2^*)$ for the partition, we have $dist(c_1^*,c_2^*)=\delta$.
\label{prop_equal_center_dist}
\end{prop}
\textbf{Proof.} We proceed by contradiction. Suppose that for an optimal solution $(D^*_1,D^*_2)$, $dist(c_1^*,c_2^*)<\delta$. So, at least one of the centers for example $c^*_1$ should be different from $c^{**}_1$ and lies inside the region between the two perpendicular lines from $c_1^{**}$ and $c_2^{**}$ on $line(c^{**}_1,c^{**}_2)$ (the line passing $c_1^{**}$ and $c_2^{**}$). This is because if both $c_1^*$ and $c_2^*$ are outside this region, the distance between them can't be less than $\delta$. If $c^*_1$ is not on $\partial \mathcal{F}(P_1)$, then we can slightly move $c^*_1$ toward its farthest point, reducing the radius of $D^*_1$ while not violating the PCC which contradicts best optimality. If $c_1^*$ is on $\partial \mathcal{F}(P_1)$, by Proposition \ref{prop_weight_change}, any point on the interior of the path from $c_1^{**}$ to $c_1^*$ on $\partial \mathcal{F}(P_1)$, covers $P_1$ with radius smaller than $r(D^*_1)$. Since $dist(c^*_1,c^*_2)<\delta$, any point on this path sufficiently close to $c_1^*$ will not violate the PCC. This contradicts the fact that $(D^*_1,D^*_2)$ is a BOS.$\hfill\square$\\\\
Let $H_1(r)$ and $H_2(r)$ be the intersection hulls of $P_1$ and $P_2$ with radius $r$. Note that the smallest radii for which the intersection hulls of $P_1$ and $P_2$ are nonempty are $r(D_1^{**})$ and $r(D_2^{**})$ respectively. Let's denote them by $r_1^0$ and $r_2^0$ (in fact $H_1(r_1^0)=c_1^{**}$ and $H_2(r_2^0)=c_2^{**}$). If $r\geq r_1^0$, for any point $q\in H_1(r)$, $disk(q,r)$(the disk with center $q$ and radius $r$) covers $P_1$ (we have a similar statement for $P_2$ and $r_2^0$). Based on this property, the problem turns to find a best optimal pair of radii $(r^*_1,r^*_2)$ (the bigger radius is minimum and the smaller radius is minimum among all such pairs) such that the distance between $H_1(r^*_1)$ and $H_2(r^*_2)$ is exactly $\delta$. Also, we call the maximum radius of optimal pair(s) the \textit{optimal cost} and denote it by $r^*$. The idea to find a best optimal pair is first try to find the optimal cost $r^*$ and then, fix one of the intersection hulls at radius $r^*$ and find minimum possible radius for the other hull. So here, we focus on finding the optimal cost.\par
In order to find the optimal cost, we impose the constraint that the disks are congruent (radii of both hulls should be equal). Imposing this constraint makes the problem easier while it does not change the optimal cost. In order to solve the problem for congruent disks, we can grow the intersection hulls of the points at each part of the partition to see when the distance between them becomes $\delta$. We first build $\mathcal{F}(P_1)$ and $\mathcal{F}(P_2)$ which can be done in $O(n\log n)$ time. In order to prevent structural changes when we grow the intersection hulls, we apply a binary search (repeatedly find the median and discard half of the values) on the set of weights of the farthest-point Voronoi diagrams of both sides to obtain an interval $I^*=(i_0,i_1)$ such that $r^*\in I^*$ and for each vertex $v$ of the diagrams, $w(v)\notin I^*$ (the weight of a point $x$ in a cell of farthest-point Voronoi diagram denoted by $w(x)$ is the distance between $x$ and the site of the cell containing it. See Appendix A for details). At each step of the binary search, when we test a weight $w$, we use the algorithm of \cite{DST} to compute the distance between the two intersection hulls at radius $w$ to see whether their distance is smaller, equal or greater than $\delta$. Note that because the intersection hulls are convex, this step can be done in $O(\log n)$ time according to \cite{DST} (we don't need to explicitely build the intersection hulls because their vertices are along the edges of the farthest-point Voronoi diagrams). 
\begin{obs}
For any index $i$ and any $r\in I^*$, the endpoints of the $i^{th}$-element of $Seq(H_1(r))$ and $Seq(H_1(i_0))$ (resp. $Seq(H_2(r))$ and $Seq(H_2(i_0))$) lie on same arms of $H_1(i_0)$(resp. $H_2(i_0)$).
\end{obs}
In other words, when $r$ varies from $i_0$ to $i_1$, no arc in intersection hulls will be emerged or vanished. Let's denote the arms of $H_1(i_0)$ by $A_1$ and call the partition induced by $H_1(i_0)\cup A_1$ the \textit{$A_1$-partition} of the plane. Now, we discuss how to find the optimal cost for the partition. Suppose that we have not found $r^*$ during the binary search (otherwise we are done). So, $dist(H_1(i_0),H_2(i_0))>\delta$ and $dist(H_1(i_1),H_2(i_1))<\delta$. Let $Seq(H_1(i_0)) = (X_1,\dots,X_u)$ and $Seq(H_2(i_0))=(Y_1,\dots,Y_{u'})$. We also label each region of the $A_1$-partition bounded by two neighbour arms by the name of the arc it contains. Each arm in $A_1$ can intersect $\partial H_2(i_0)$ in at most two points. Consider an arm in $A_1$ with endpoint $a$ and an intersection point $x$ with $\partial H_2(i_0)$. We call this intersection point a \textit{first intersection point} if $ax$ does not intersect the interior of $H_2(i_0)$. \par
\begin{figure}
\begin{center}
\includegraphics[scale=0.7]{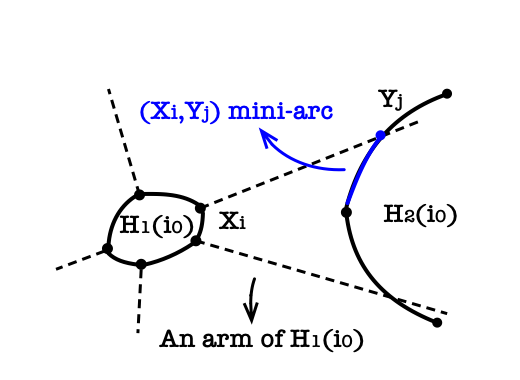}
\end{center}
\caption{A mini-arc for two intersection hulls}
\label{fig_mini_arc}
\end{figure}

Let $B$ be the set of all first intersection points of the arms in $A_1$ and $\partial H_2(i_0)$. Note that $H_2(i_0)$ is convex and the arms around $H_1(i_0)$ diverges from each other. Also, we already have the order of the arms around $H_1(i_0)$ induced by $\mathcal{F}(P_1)$. In order to compute $B$, consider the counter-clockwise order on the arms of $A_1$ starting from the arm with the lowest slope (can be negative) and compute their first intersection points with $H_2(i_0)$ in order. An important point here is that if a vertex of $H_2(i_0)$ lies on the right side of an arm $\vec{a}$ (the direction is from its endpoint), it will be on the right side of any arm after $\vec{a}$. This proprty implies that the cost of computing $B$ is linear (see Figure \ref{fig_mini_arc}).\par
\begin{figure}
\begin{center}
\includegraphics[scale=0.35]{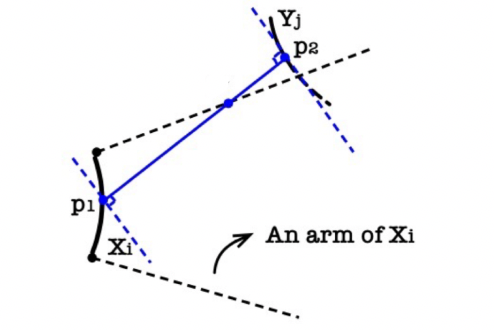}
\end{center}
\caption{Connecting line of two points from two arcs which do not make a mini-arc intersects an arm}
\label{fig_min-arcs_proof}
\end{figure}

      
Now, consider the partition induced by $B$ and the vertices of $\partial H_2(i_0)$ on $\partial H_2(i_0)$. We call each region of this partition a \textit{mini-arc} on $\partial H_2(i_0)$ (see Figure \ref{fig_mini_arc}). 
We assign a label $(X_i,Y_j)$ to each mini-arc of $\partial H_2(i_0)$ where $X_i$ is the label of the mini-arc in the $A_1$-partition and $Y_j$ is the label the arc of $H_2(i_0)$ containing the mini-arc. Note that the number of such $(X_i,Y_j)$ labels are linear (because we have a linear number of mini-arcs). 
\begin{prop}
The labels of the two arcs containing two closest points between $H_1(r^*)$ and $H_2(r^*)$ corresponds to the label of one of the mini-arcs.
\label{prop_closest_points_hull}
\end{prop}
\textbf{Proof.} Let $p_1$ and $p_2$ be two points on $H_1(r^*)$ and $H_2(r^*)$ respectively with distance $\delta$. First, we observe that the perpendicular lines on $line(p_1,p_2)$ from $p_1$ and $p_2$ should not intersect the interior of $H_1(r^*)$ and $H_2(r^*)$ (otherwise it contradicts the optimallity of $r^*$). Suppose that $p_1$ and $p_2$ are lie on two arcs $X_i$ and $Y_j$ respectively (we consider the names of the arcs in $H_1(r^*)$ and $H_2(r^*)$ the same as the label of their corresponding regions in $H_1(i_0)$ and $H_2(i_0)$ respectively). If $(X_i,Y_j)$ is not a label of a mini-arc, $p_1p_2$ should intersect an arm of $X_i$. But in this situation, the perpendicular line on $line(p_1,p_2)$ from $p_1$ should intersect the interior of $H_1(r^*)$ (because of convexity of $H_1(r^*)$) which is contradiction. See Figure \ref{fig_min-arcs_proof}. $\hfill\square$\\\\
According to Proposition \ref{prop_closest_points_hull}, we can compute $r^*$ as follows: we consider each label $(X_i,Y_j)$ of the mini-arcs and compute the value $r_{i,j}$ for which the distance between the two arcs $X_i$ and $Y_j$ becomes $\delta$ as they propagate between their bounding arms. Note that computing each $r_{i,j}$ can be done in a constant time. Now, $r^*$ is the minimum value among all $r_{i,j}s$.\par
The next step is obtaining a BOS for the partition after computing $r^*$. As we said earlier, in order to do this, we first assume that the determining disk covers $P_1$ and obtain the minimum possible radius $r_1'$ for $H_2$ which makes the distance between $H_1(r^*)$ and $H_2(r'_1)$ exactly $\delta$. This can be done in a similar way to how we obtained $r^*$. Then we obtain $r_2'$ similarly by assuming that the determining disk covers $P_2$. By comparing the results, we pick the one with smaller non-determining disk which is in fact a BOS for the partition. So, the total time we need to compute a BOS is $O(n\log n)$.
\section{Obtaining a BOS for the PCTC Problem}
% Another geometric class of objects which is closely related to farthest-point Voronoi diagrams is \textit{intersection hull}s. For a given set of points $A$, the \textit{intersection hull} of $A$ with respect to a radius $r$ is defined as the intersection of all disks with radius $r$ and center in $A$. We call $r$ the \textit{radius} of the hull. Let $H_A(r)$ be the intersection hull of $A$ at radius $r$. Then, the boundary of $H_A(r)$ is consisted of a set of arcs with radius $r$. Each arc completely lies in a cell of $\mathcal{F}(A)$ (farthest-point Voronoi diagram of $A$) with each endpoint on an edge of $\partial \mathcal{F}(A)$ (boundary of the farthest-point Voronoi diagram of $A$). Suppose that $C_i$ and $C_j$ are two neighbour arcs of $H_A(r)$ with centers $a_i$ and $a_j$ respectively. Also, let $x$ be their common endpoint. Then, the perpendicular bisector of $a_i$ and $a_j$ passes $x$. We call the half-line along this bisector from $x$ that does not intersect the interior of $H_A(r)$ an \textit{arm} of $H_A(r)$. See Figure \ref{fig1} (a) for an illustration.
%\begin{figure}
%\begin{center}
%\includegraphics[scale=0.3]{Fig1v2.png}
%\end{center}
%\caption{a) The farthest-point Voronoi diagram of four points and an intersection hull $H(r)$ for it. Also, it shows the arm of $H(r)$ corresponding to $(a_3,a_4)$. b) Construction of $\pi(r)$.}
%\label{fig1}
%\end{figure}
%\vspace{-0.4cm} 
%\section {Computing a BOS for the Problem}
%\vspace{-0.2cm}
We denote the optimal cost for the PCTC problem by $r^*$ and a BOS for the problem by $(D^*_1,D^*_2)$ with centers $(c^*_1,c^*_2)$ respectively. We can assume that $c^*_1$ and $c^*_2$ lie on the $x$-axis and $c^*_1$ is on the left side of $c^*_2$. In \cite{SH}, Sharir broke the $2$-center decision problem (given a parameter $r$ determine whether it is possible to cover the points with two disks of radius $r$) into three cases -far distant, distant and nearby- with respect to the given parameter $r$. He showed that providing separate algorithms for these cases will reduce the overall time complexity to solve the decision problem. Although our problem is an optimization problem and the parameter $r^*$ is unknown, we will show that breaking the PCTC problem into the same cases will simplify our algorithm and reduce the overall time complexity. So, our algorithm separately considers each of the following three assumptions about $(D^*_1,D^*_2)$.   
\begin{enumerate}
\item
Nearby: $dist(c^*_1,c^*_2)\leq r^*$.
\item 
Distant: $r^*< dist(c^*_1,c^*_2)\leq 3r^*$. 
\item
Far distant: $dist(c^*_1,c^*_2)>3r^*$.
\end{enumerate}
Denote the smallest cost we can get having the nearby, distant and far distant assumptions by $r^{NA}$, $r^{DA}$ and $r^{FA}$ respectively. We also use the same notation for a BOS and their corresponding centers having each assumption. So, we can obtain $(D^*_1,D^*_2)$ by comparing $(D^{NA}_1,D^{NA}_2)$, $(D^{DA}_1,D^{DA}_2)$ and $(D^{FA}_1,D^{FA}_2)$ (note that these solutions may not exist or satisfy their corresponding case conditions. For example, $dist(c^{NA}_1,c^{NA}_2)$ might be greater than $r^{NA}$ but if $(D^*_1,D^*_2)$ satisfies the nearby case, then $r^{NA}=r^*$ and $(D^{NA}_1,D^{NA}_2)$ would be a BOS for the problem and we have $dist(c^{NA}_1,c^{NA}_2)\leq r^{NA}=r^*$). Henceforth, while studing each of the cases, when we say BOS, we mean a best solution we can get having the corresponding case assumption. Given two points $x$ and $y$ in the plane, we denote the line passing from $x$ and $y$ by $line(x,y)$. The direction of this line is considered from $x$ to $y$. Also, we denote the half-line from $x$ passing $y$ by $half\text{-}line(x,y)$ and the line segment with end points $x$ and $y$ by $seg(x,y)$. 
\section{Computing a BOS in the Nearby Case}
\begin{figure}
\begin{center}
\includegraphics[scale=0.35]{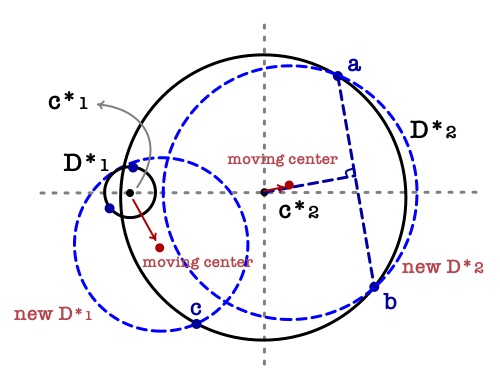}
\end{center}
\caption{Enlarging the non-determining disk $D^*_1$ to cover one of the dominating points of $D^*_2$ and get a better solution.}
\label{fig_constant_factor}
\end{figure}

First, we can see that if $(D^*_1,D^*_2)\leq r^*$, then there is an optimal partition $R^*$ (may not be unique) such that $(D^*_1,D^*_2)$ is a BOS of $R^*$. In fact, such a partition can be obtained by considering a point in $D^*_1\cap D^*_2$ and two half-lines from it passing the intersection points of $\partial D^*_1$ (boundary of $D^*_1$) and $\partial D^*_2$. In this section, when we say the dominating points of $(D^*_1,D^*_2)$, we mean its dominating points with respect to $R^*$. Without loss of generality, we can assume that $D^*_2$ is the determining disk. We first compute the $convex\text{-}hull(P)$ and scale the problem such that it fits in a unit square (multiple both $x$ and $y$ coordinates of the points by the greatest constant such that the convex hull remains inside the square). This step can be done in $O(n\log n)$ time. Note that the scaling will not change the solutions. 
\begin{prop}
If $(D^*_1,D^*_2)\leq r^*$, then the area of $D^*_1\cap D^*_2$ must be greater than a constant factor of the area of $D^*_2$ (the determining disk).
\label{prop_int_area}
\end{prop}
\textbf{Proof.} We proceed by contradiction. Suppose that such a factor does not exist. This means that we can build a problem instance such that it has a BOS $(D^*_1,D^*_2)$ in which the radius of the non-determinig disk ($D^*_1$) becomes infinitely small (because of the nearby assumption and scaling). So, $D^*_1$ should have at least one dominating point that is not covered by $D^*_2$. Because the radius of $D^*_1$ is infinitely small, $\delta$ should tend to $radius(D^*_2)$ (which tends to the radius of the MED of $P$). Now, $D^*_2$ should have at least one dominating point (point $c$ in Figure \ref{fig_constant_factor}) with the $x$-coordinate less than or equal to $c^*_2$ (otherwise, we can move both $c^*_1$ and $c^*_2$ to the right and reduce the radius of $D^*_2$ which determines the cost). In this configuration, we can enlarge $D^*_1$ by moving $c^*_1$ toward this dominating point of $D^*_2$ while satisfying the PCC in order to cover it and release it from $D^*_2$ ($D^*_1$ does not lose any of its own points and its radius still remains less than the radius of $D^*_2$). Now, we can reduce the radius of $D^*_2$ which contradicts the optimallity of $(D^*_1,D^*_2)$ (see Figure \ref{fig_constant_factor}).$\hfill\square$\\\\
\begin{prop}
$D^*_1$ (similarly $D^*_2$) should have a pair of dominating points such that:
\begin{enumerate}
\item
They lie on different sides of $line(c^*_1,c^*_2)$.
\item
Their connecting segment does not intersect $seg(c^*_1,c^*_2)$.
\end{enumerate}
\label{prop_dif_int}
\label{prop_dom_dif_sides}
\end{prop}
See Appendix B for the proof. Considering the four dominating points in the above proposition, we can say that $D^*_1\cap D^*_2$ should cover at least a constant factor of the area of $convex\text{-}hull(P)$. Furthermore, $D^*_1\cap D^*_2\cap convex\text{-}hull(P)$ is convex because it is the intersection of convex objects. So, we can build a constant size set of points $\mathcal{M}$ uniformly distributed on $convex\text{-}hull(P)$ such that (assuming  $dist(c^*_1,c^*_2)\leq r^*$) for at least one point $\hat{m}\in \mathcal{M}$, $\hat{m}\in D^*_1\cap D^*_2\cap convex\text{-}hull(P)$. Because $\hat{m}$ is unknown, for each $m\in \mathcal{M}$, we build a BOS $(D_1^m,D_2^m)$ assuming $m\in D_1^*\cap D_2^*$ and finally pick a best solution in $\{(D_1^m,D_2^m)\;:\;m\in \mathcal{M}\}$ and set it as $(D^{NA}_1,D^{NA}_2)$. Based on this idea, we present our algorithm to find $(D^m_1,D^m_2)$ for a given point $m\in convex\text{-}hull(P)$. \par
Let $\mathcal{X}$ be a set of $360$ directed lines (each line has a positive direction) passing through $m$ such that the angle between each directed line and its neighbour lines is $1^\circ$. Now, there should be a directed line in $\mathcal{X}$ such that its angel with $line(c^*_1,c^*_2)$ is at most $1^\circ$ and $c^*_1$ lies on the negative side of $c^*_2$ on the line (note that $D^*_2$ is the determining disk according to our assumption). We call this directed line the \textit{correct directed line} which is unknown. So, we assume each line $l\in \mathcal{X}$ as the correct directed line and compute a BOS $(D^{m,l}_1,D^{m,l}_2)$ having this assumption and finally pick the best one as $(D^m_1,D^m_2)$. \par
\begin{figure}[h]
\begin{center}
\includegraphics[scale=0.14]{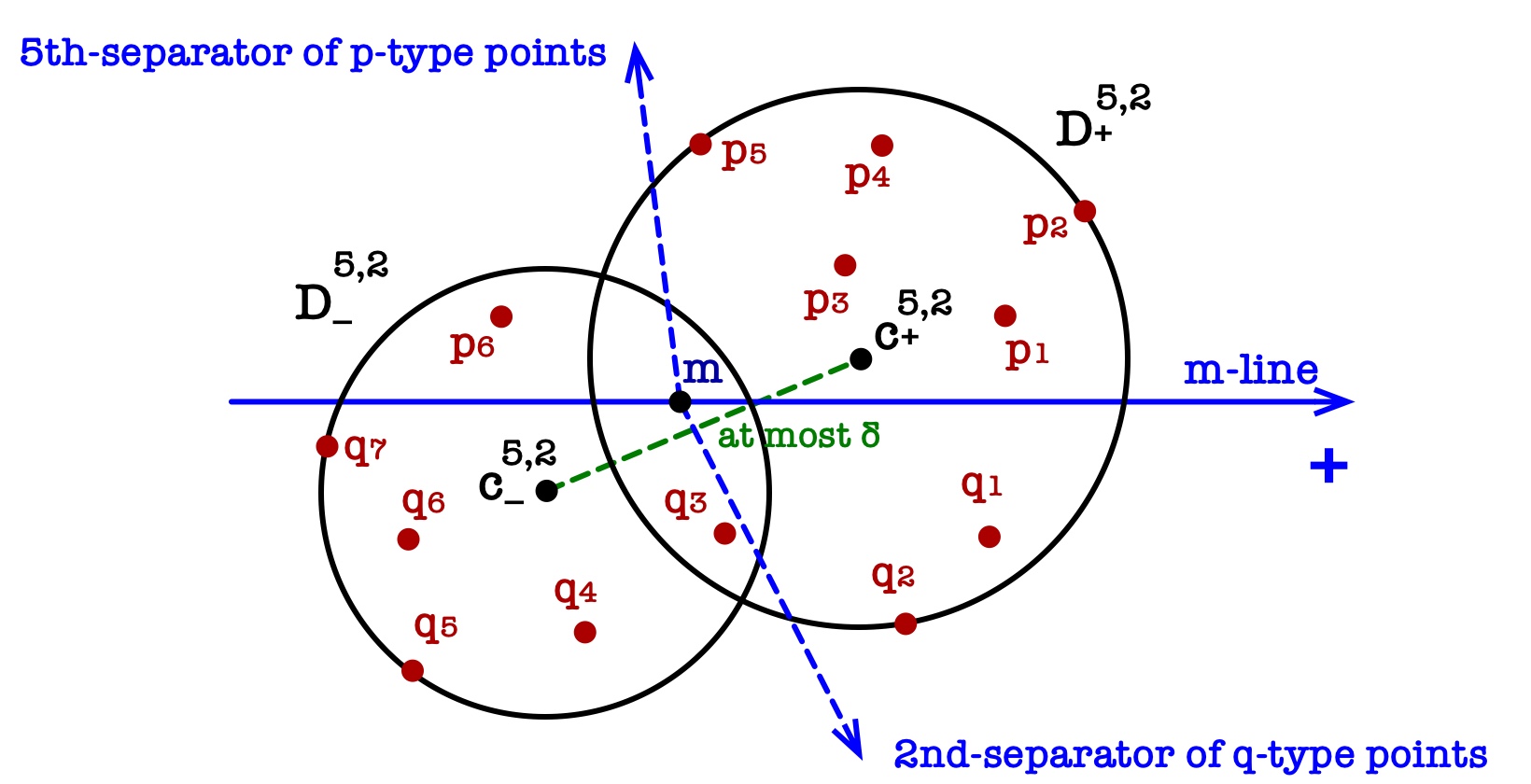}
\end{center}
\caption{An example $(i,j)$-partition of a set of points and a BOS for the partition.}
\label{fig_partition}
\end{figure}

So, assume that a directed line $l\in \mathcal{X}$ called the $m$-line is given. Here we explain how to compute $(D^{m,l}_1,D^{m,l}_2)$. The $m$-line divides the points of $P$ into two disjoint sets one on the right side and the other on the left side of the $m$-line. We sort these sets according to the polar angles of their points (from $m$) with respect to the positive direction of the $m$-line. These angles should lie between $-180^\circ$ and $180^\circ$ and we sort them by increasing magnitude (see Figure \ref{fig_partition} for an illustration). Based on these orders, we denote the two sequences of points on the left and right side of the $m$-line by $(p_1,\dots,p_{n'})$ and $(q_1,\dots,q_{n''})$ respectively. We call a point $p\text{-}type$ (resp. $q\text{-}type$) if it is in the first(resp. second) sequence. We also call a half-line from $m$ that separates $\{p_1,\dots,p_i\}$ from $\{p_{i+1},\dots,p_{n'}\}$ an $i^{th}$-\textit{separator} of the $p$-type points. A $j^{th}$-separator of $q$-type points is defined similarly (we assume that the $0^{th}$ and $n'^{th}$(resp. $n''^{th}$) separators have the entire $p$-type(resp. $q$-type) points in one side). The $i^{th}$ and $j^{th}$ separators of the $p$-type and $q$-type points partition the plane into two parts. We call this partition the $(i,j)$-\textit{partition} of the plane. One part of this partition contains the positive direction of the $m$-line which we call it the \textit{positive side} of the partition and we call the other part the \textit{negative side} of the partition.
\begin{obs}
If $dist(c^*_1,c^*_2)\leq r^*$, $m=\hat{m}$ and the $m$-line is correct, then an $(i,j)$-partition can be considered as $R^*$ and $(D^*_1,D^*_2)$ is its BOS.  
\end{obs} 
Note that in the above observation, the two separators from $m$ passing the intersection points of $D^*_1$ and $D^*_2$ give us the desired $(i,j)$-partition. We denote the set of points in the positive and negative sides of the partition by $P^{i,j}_+$ and $P^{i,j}_-$ respectively. Based on our algorithm for restricted PCTC problem, a BOS for an $(i,j)$-partition can be computed in $O(n\log n)$ time. Let $(D^{i,j}_-,D^{i,j}_+)$ (with centers $(c^{i,j}_-,c^{i,j}_+)$ respectively) be the output of this algorithm for the $(i,j)$-partition (see Figure \ref{fig_partition} for an example). We refer to the first(resp. second) disk the \textit{negative disk}(resp. \textit{positive disk}) of the partition. A naive approach to obtain $(D^*_1,D^*_2)$ is to apply our restricted PCTC problem algorithm to each of the $(i,j)$-partitions and pick the best one. This will give us an $O(n^3\log n)$ time complexity as there are quadratic partitions. In the following we show how we can get $(D^{m,l}_1,D^{m,l}_2)$ by evaluating a sub-quadratic number of partitions. The idea is first computing $r^{m,l}$ which is the best cost we can get assuming $m$ and $l$ are correct. Then, we use it to compute $(D^{m,l}_1,D^{m,l}_2)$. 
\vspace{-.4cm}
\subsection{Computing $r^{m,l}$}
Let's define $M^+$ as a $(n'+1)\times (n''+1)$ matrix whose $[i,j]$-element ($0\leq i\leq n'$ and $0\leq j\leq n''$) is $radius(D^{i,j}_+)$. We call $M^+[i,j]$ \textit{non-critical} if $D^{i,j}_+$ is the MED of $P^{i,j}_+$. Otherwise, we call it \textit{critical}. We call $M^+[i,j]$ a \textit{valid} element if $M^+[i,j]\geq radius(D^{i,j}_-)$ and we call it \textit{non-valid} otherwise. Because we assumed that $l$ is correct, we can assume that positive disks determine $r^{m,l}$. This means that $r^{m,l}$ is indeed the minimum valid element of $M^+$.\par
\begin{prop} For any $0\leq i\leq n'$ and $0\leq j\leq n''$, we have:
\begin{enumerate}
\item
If $M^+[i,j]$ is non-critical, then $M^+[i',j']\geq M^+[i,j]$ for all $i'\geq i$ and $j'\geq j$.
\item
If $M^+[i,j]>radius(D^{i,j}_-)$, $M^+[i,j]$ is non-critical.
\item
If $M^+[i,j]$ is valid and critical, then $M^+[i,j]=radius(D^{i,j}_-)$ and $dist(c^{i,j}_-,c^{i,j}_+)=\delta$.
\end{enumerate}
\label{prop_123_matrix}
\end{prop}
Briefly, case 1 is clear and if each of the cases 2 or 3 is not true, by moving the centers we can get a better solution. See Appendix B for details. We search $M^+$ to find $r^{m,l}$ as follows: we maintain a set of \textit{candidate values}. During the search, when we evaluate an element $M^+[i,j]$(computing $(D^{i,j}_-,D^{i,j}_+)$ and its dominating points), if $M^+[i,j]$ is valid, we add it to the candidate values and finally we set $r^{m,l}$ as the minimum candidate value. \par
In order to search $M^+$, we maintain two variables $I$ and $J$ where $I$ stores the index of the current row that we are searching and $J$ stores the column index for which we can discard any column with index greater than that. Initially, we set $I=0$, $J=n''$ ($n''$ is the number of columns of $M^+$). We search the $I^{th}$-row by evaluating its elements backward starting from its $J^{th}$-element (if $J=-1$, the matrix search is done) toward its first element. Because we are looking for a minimum valid element of the matrix, we can use Proposition~\ref{prop_123_matrix} to improve our search as follows: during the traversal of the row, if $M^+[I,j]$ is valid and non-critical, we set $J=j-1$ (because $D^{I,j}_+$ is the MED of $P^{I,j}_+$, when we add more points to the positive side we can't get a smaller positive disk). We finish traversing the row and increase $I$ by one if either the row is exhausted or we reach an index $j$ such that $M^+[I,j]$ becomes non-valid. Note that in this case, $D^{I,j}_-$ is the MED of $P^{I,j}_-$ (similar to Proposition \ref{prop_123_matrix} part 3). Here we might have a valid element on some index $j'<j$ but, the cost of this solution can not be less than $radius(D^{I,j}_-)$ (we add points to the negative side as we move left wise on a row). In order to make sure that we will count such costs in our algorithm, we can add $radius(D^{I,j}_-)$ to the candidate set of the directed line in $\mathcal{X}$ with the opposite direction of the current $m$-line.\par
We continue this procedure until no element is left. Note that when none of $D^{i,j}_-$ and $D^{i,j}_+$ are a MED, we can't discard any element from the matrix because it is possible that when we move a point from one side to the other, the radii of both disks become greater or smaller while they remain equal (this situation can happen because of the PCC). So the number of evaluations in the above schema might be still quadratic. Next, we explain how to fix this problem.
\begin{prop}
If $M^+[i,j]$ is valid-critical and $q_j$ is not a dominating point of $D^{i,j}_+$, then $M^+[i,j-1]\geq M^+[i,j]$.
\label{prop_discard_critical}
\end{prop}
\textbf{Proof.} Because $D^{i,j}_+$ is not a MED, its center can't get closer to its farthest points in $P^{i,j}_+$ (dominating points of $D^{i,j}_+$) namely $d_1$ and $d_2$ because of the PCC. Now, by adding $q_j$ to $P^{i,j}_-$, $D^{i,j-1}_-$ needs to cover more points. If its radius gets bigger, the proposition follows. Otherwise, according to the fact that $q_j$ is not a dominating point of $D^{i,j}_+$, it is not possible to put $c^{i,j-1}_-$ on a place such that allow $c^{i,j-1}_+$ to get closer to $d_1$ and $d_2$ due to best optimality of $(D^{i,j}_-,D^{i,j}_+).\hfill\square$\\\\ Note that in this proposition, if $q_j$ does not become a dominating point of $D^{i,j-1}_-$, then $M^+[i,j-1]=M^+[i,j]$. A similar statement is also correct for two consecutive valid-critical elements in a column. Based on the above proposition, we can improve our matrix search as follows: while traversing a row(left wise), when we hit a valid-critical element $M^+[i,j]$, if both dominating points of $D^{i,j}_+$ are $p$-type, we discard the rest of the row (because by traversing a row, only $q$-type points will move to the other part of the partition) and continue the search on the next row. Similarly, if both dominating points of $D^{i,j}_-$ are $q$-type, we can discard the rest of its column. Otherwise, we jump to the first (largest index) element of the row for which a $q$-type dominating point of $D^{i,j}_+$ moves to the negative side and discard all the elements in between (because of Proposition \ref{prop_discard_critical}). Similarly, we discard the portion of the rest of the column of $M^+[i,j]$ with row index smaller than the index of the $p$-type dominating point(s) of $D^{i,j}_-$(applying the column version of Proposition \ref{prop_discard_critical}). \par
When we evaluate a valid-critical element $M^+[i,j]$, if we didn't discard the entire rest of its row or column, we \textit{mark} the portion of its column that is not discarded after the evaluation of $M^+[i,j]$. Now, when we traverse the rows, we ignore and jump discarded and marked elements. Specially, if after evaluating an element $M^+[i,j]$, the largest index of the $q$-type dominating point of $D^{i,j}_+$ is $j'$ and $M^+[i,j']$ is marked, we continue searching from the first(biggest index) unmarked or undiscarded element of the $i^{th}$-row after $M^+[i,j']$. Applying this marking schema in the matrix search will guarantee that the number of evaluations is linear. The problem of our matrix search with marking schema is that we may mark the minimum valid element of $M^+$ and so get an incorrect $r^{m,l}$. In the rest, we will show how to overcome this problem.\par  
We call the above matrix search \textit{initial search} of $M^+$ from \textit{top-right}. Another way of searching $M^+$ is starting the search from $M^+[n',0]$ (the first element of the last row). But this time, instead of traversing the rows from right to left, we traverse the columns from bottom to top. The way we search the matrix is exactly symmetrical to the top-right search but here we mark sub-rows instead of sub-columns. We call this matrix search the initial search of $M^+$ from bottom-left. After performing two initial searches on $M^+$ one from the top-right and one from the bottom-left, still there might be some elements that are marked in both initial searches. We call these elements as \textit{doubly-marked} elements. The next theorem enables us to search the doubly-marked elements in an efficient way which leads us to find $r^{m,l}$. Lets denote the doubly-marked elements of $M^+$ by $Doubly\text{-}Marked(M^+)$. 
\begin{theorem}
By evaluating a doubly-marked element $M^+[i,j]$, we can discard one of the following sub-rows or sub-columns of $M^+$:
\begin{enumerate}
\item
Elements above $[i,j]$ ($M^+[i',j]$ with $i'\leq i$).
\item
Elements below $[i,j]$ ($M^+[i',j]$ with $i'\geq i$).
\item
Elements in front of $[i,j]$ ($M^+[i,j']$ where $j'\geq j$).
\end{enumerate}
\label{main_theorem}
\end{theorem}
Suppose that $M^+[\hat{i},\bar{j}]$ is a given doubly-marked element which is marked when we evaluate $M^+[\bar{i},\bar{j}]$ and $M^+[\hat{i},\hat{j}]$ in the initial top-right and bottom-left search respectively. When we evaluate $M^+[\hat{i},\bar{j}]$, we get $D^{\hat{i},\bar{j}}_+$ and $D^{\hat{i},\bar{j}}_-$ and their dominating points. For the sake of simplicity, let's denote the first disk by $D'_+$ and the second disk by $D'_-$. If $D'_-$ is MED, then either $radius(D'_-)\geq radius(D'_+)$  or $radius(D'_-)< radius(D'_+)$. In the former, case 1 in Theorem \ref{main_theorem} happens and in the latter, $D'_+$ should be MED (otherwise we can reduce its cost and the solution can't be optimal) and so cases 2 and 3 of the theorem happen. We have a similar argument when $D'_+$ is a MED. So, the only left case is when none of the disks is MED. Note that in this case each of $D'_+$ and $D'_-$ has exactly two dominating points. Let $h_1, h_2$ be the dominating points of $D'_+$ and $h'_1, h'_2$ be the dominating points of $D'_-$. If both $h'_1$ and $h'_2$ are $p$-type, case 3 happens (when we traverse the $\hat{i}^{th}$-row from left to right, we only add $q$-type points to the positive side). Also, if they are both $q$-type, case 2 happens. The bottleneck of proving Theorem~\ref{main_theorem} is when $h'_1$ and $h'_2$ have different types. In order to prove Theorem \ref{main_theorem} in this special case, we use two key properties. First $M^+[\hat{i},\bar{j}]$ should be doubly-marked and second, $m$ should be inside the convex hull of the points. We leave this proof Appendix C and in the rest, we focus on how to use Theorem \ref{main_theorem} to search $Doubly\text{-}Marked(M^+)$ efficiently.
\subsection{Searching the Doubly-Marked Elements}
For simplicity, we assume that $n'=2^g-1$ for some integer value $g>1$ (so the number of rows is a power of $2$). We define the $k^{th}$-division of $M^+$ as the sub-matrix consisting of the rows from $n'/2^k$ to $n'/2^{k-1}-1$ (we search the first row independently by evaluating all of its doubly-marked elements). We search the divisions of $M^+$ in order from its first division. Let's denote the $k^{th}$-division sub-matrix by $DIV_k$. Here, we explain how to search $DIV_k$. Let $I$ and $J$ be the row and column indices (with respect to $DIV_k$) of the element that we are processing at each time. Initially, we have $I=J=1$ (the first row and column of $DIV_k$). We evaluate the non-discarded elements of the $I^{th}$-row from left to right starting from the column index $J$. If the result of an evaluation is case 1 or 2 in Theorem \ref{main_theorem}, we discard the corresponding portion of $M^+$ (in all divisions) and increase $J$ by one. But if case 3 happens, we go to the next row and increase $I$ (we always move rightwise). After we proceed with all divisions, some elements might left unevaluated and undiscarded in each division due to the occurrence of case 1. We recursively perform the entire above process on these unevaluated elements in each division until all elements are either discarded or evaluated. So, if only doubly-marked elements remained in $M^+$ (we have discarded all other elements in the two initial searches), then the procedure SEARCH-DM($M^+$) in Algorithm 1 will give us a minimum valid element of $M^+$.
\begin{algorithm}[h]
\caption{SEARCH-DM($M$)}
\begin{algorithmic}[1]
\State Let $M$ be a $n\times m$ matrix.
\State Split $M$ into $\log n$ divisions $\{DIV_1,\dots,DIV_{\log n}\}$ .
\For {$k=1,\dots,\log n$} // We search the divisions in order.
\State Set $I=J=1$.
\Repeat
\State Evaluate $DIV_k[I,J]$ and discard the portion of $M$ according to Theorem \ref{main_theorem}.
\If {(case 1 or 2 happens) \textbf{and} $J<m$} $J=J+1$.
\Else
\State $I=I+1$.
\EndIf
\Until {$I>n/2^k$} // number of rows in $DIV_k$.
%\State Set $M'$ as the undiscarded and unevaluated elements of $DIV_k$.
\State SEARCH-DM($DIV_k$) if $DIV_k$ has unevaluated/undiscarded element.
\EndFor
\State Evaluate all non-discarded elements of the first row of $M$. // Until the case 3 happens.
\end{algorithmic}
\end{algorithm}
\begin{theorem}
SEARCH-DM($M^+$) evaluates $O(n\log n)$ elements of $M^+$. 
\end{theorem}
\textbf{Proof.} First, if only cases 2 and 3 happen in the algorithm, then we don't need the recursion part and so the total number of evaluations becomes $O(n\log n)$ (in each iteration of searching $DIV_k$ either $I$ or $J$ would be increased). Now, suppose that any of the cases 1, 2, or 3 can happen. Note that the number of case 3s in all divisions of a same recursion level (the original $\log n$ divisions has recursion level zero and the level of the divisions in the recursion part of the algorithm is defined based on their appearance in the recursion tree) is at most $n$ because two divisions of a same level has disjoint rows. Because we have $O(\log n)$ levels, the total number of case 3 evaluations is $O(n\log n)$. Now, if after the evaluation of some $DIV_k[i,j]$, case 1 happens, we can't discard any new element from $DIV_k$ but all the elements above $DIV_k[i,j]$ in $M$ should be discarded. This means that while searching each of the divisions $DIV_{k+1},\dots,DIV_{\log n}$ and the first row, we don't need to evaluate the $j^{th}$-column. On the other hand, $DIV_k$ has $\log (n/2^{k})=\log n-k$ divisions and a row. Each of these divisions can have at most one cases 1 or 2 in the $j^{th}$-column. So, we can have a correspondence between the extra cases 1 and 2 evaluations in searching the divisions and the first row of $DIV_k$ (not its recursion part) and the matrix elements that we didn't evaluate in $DIV_{k+1},\dots,DIV_{\log n}$. So, the total number of evaluations would remain $O(n\log n)$.$\hfill\square$\\\\
Note that in a constant time, we can check whether an element is discarded or not. Because in each recursion level, the divisions are disjoint, at each level we check each element of $M$ at most once and because we have $O(\log n)$ levels, the total cost of matrix element checking would be $O(n^2\log n)$. On the other hand, our algorithm to solve the restricted PCTC problem costs $O(n\log n)$, if we directly use it to evaluate matrix elements, the total time complexity of SEARCH-DM($M^+$) becomes $O(n^2\log^2 n)$. As we mentioned in Section 2, the bottleneck of solving the restricted PCTC problem is computing the farthest-point Voronoi diagram of each part of the partition which costs $O(n\log n)$. So, if we can reduce this cost by performing a preprocessing step, we can reduce the overall time complexity of SEARCH-DM($M^+$). In order to speed up matrix element evaluation, we use the following lemma from \cite{DNK}:
\begin{lemma}\cite{DNK}
If $X$ and $Y$ are arbitrary sets of points in the plane, then $\mathcal{F}(X\cup Y)$ can be constructed from $\mathcal{F}(X)$ and $\mathcal{F}(Y)$ in $O(|X|+|Y|)$ time ($\mathcal{F}(X)$ represents the farthest-point Voronoi diagram of $X$).
\label{L1}
\end{lemma}
\textbf{The preprocessing step:} Let $(X^+_i,X^-_i)$ (resp. $(Y^+_j,Y^-_j)$) be the partition of the $p$-type (resp. $q$-type) points induced by the $i^{th}$-separator (resp. $j^{th}$-separator). In the preprocessing phase, we compute the farthest-point Voronoi diagram of all $X^+_i$, $X^-_i$, $Y^+_j$ and $Y^-_j$ for $0\leq i\leq n'$ and $0\leq j\leq n''$. This step can be done in $O(n^2)$ using Lemma \ref{L1} because as $i$ or $j$ increases or decreases by one, a point from one side would be added to the other side.\\\\
Now, we can reduce the cost of matrix element evaluation as follows: In order  to evaluate $M^+[i,j]$, we construct $\mathcal{F}(P^{i,j}_+)$(resp. $\mathcal{F}(P^{i,j}_-)$) in $O(n)$ time by applying Lemma \ref{L1} to $\mathcal{F}(X^+_i)$ and $\mathcal{F}(Y^+_j)$ (resp. $\mathcal{F}(X^-_i)$ and $\mathcal{F}(Y^-_j)$). So, the total complexity of matrix evaluation would be $O(n)$. This reduces the time complexity of SEARCH-DM($M^+$) and so the cost of finding $r^{m,l}$ to $O(n^2\log n)$.
\subsection{Obtaining $(D^{m,l}_-,D^{m,l}_+)$ having $r^{m,l}$}     
Note that we already have an initial solution that is optimal and its cost is $r^{m,l}$ (from our search for $r^{m,l}$). But, there might be another optimal solution with the same cost and a smaller non-determining disk that we discarded during the search. If this initial solution is not best optimal, then the non-determining disk of a BOS must be strictly smaller than its determining disk. So, we can assume that the positive disk of the BOS should be the MED of the points in the positive side. Consider a matrix $\bar{M}$ for which its $(i,j)^{th}$-element is the radius of the MED of the points on the positive side of the $(i,j)$-partition. We search $\bar{M}$ from the last element of its first row and traverse the rows backwards (similar to the initial top-right search).  After evaluating an element $(i,j)$ of the matrix (which can be done in linear time according to \cite{MG}), if it is bigger than $r^{m,l}$, we discard all elements $(i,j')$ of the matrix with $j'\geq j$ because they are all greater than $r^{m,l}$ and if it is less than $r^{m,l}$, we discard the elements with $i'\leq i$ because they are all less than $r^{m,l}$. But, when it is exactly $r^{m,l}$, we compute its non-determining disk using the restricted PCTC problem algorithm (costs $O(n\log n)$) and store its radius. Here, we can also discard all elements $(i',j)$ of the matrix with $i'\leq i$. This is because as we advance more left into the row, we would have more points on the negative side and so if there is any optimal solution on the left of $(i,j)$ in the row, its non-determining disk should cover more points and thus can't give us a better solution. So, by each evaluation, we discard a row or a column of the matrix which means that the total number of evaluations is linear. Therefore, the total complexity of finding a BOS given $r^{m,l}$ would be $O(n^2\log n)$. Combining it with the complexity of computing $r^{m,l}$ gives us the total time complexity $O(n^2\log n)$ to obtain $(D^{NA}_1,D^{NA}_2)$.

\section {Computing a BOS in the Far Distant and Distant Cases}
For the far distant case, we assume that $dist(c^*_1,c^*_2)>3r^*$. In this situation, the approach of Sharir's far distant case \cite{SH} for the decision $2$-center problem still works as follows: set an arbitrary point in the plane as the origin and build $360$ directed lines $\mathcal{X}$ passing from the origin such that the degree between each line and its neighbours is $1^\circ$. Then for one unknown correct line $\vec{x}_c\in\mathcal{X}$, the angle between $line(c^*_1,c^*_2)$ and $\vec{x}_c$ is at most $1^\circ$. Supose that we set $\vec{x}_c$ is the $x$-axis and sort the $x$-coordinates of the points in $P$ as a sequence $(x_1,\dots,x_n)$. Now, if we consider the set of lines $\mathcal{L}_F^{\vec{x}_c}=\{x_i\bot x_{i+1}:1\leq i<n\}$ ($x_i\bot x_{i+1}$ is the vertical line on $\vec{x}_c$ at the mid-point of $[x_i,x_{i+1}]$), at least one $l\in \mathcal{L}_F^{\vec{x}_c}$ will separate $D^*_1$ from $D^*_2$. Because $\vec{x}_c$ is unknown, we build $\mathcal{L}_F^{\vec{x}}$ for all $\vec{x}\in\mathcal{X}$ and set $\mathcal{L}_F=\bigcup_{\vec{x}\in \mathcal{X}} \mathcal{L}_F^{\vec{x}}$. Note that the number of lines in $\mathcal{L}_F$ is linear. Here, each line $l\in \mathcal{L}_F$ induces a partition on $P$. We apply our algorithm for the restricted PCTC problem to each of such partitions and set the best one as $(D^{FA}_1,D^{FA}_2)$. So, the time complexity of the far distant case would be $O(n^2\log n)$. 
\subsection{Computing a BOS for the Distant Case}
In the distant case, we assume that $r^*< dist(c_1^*,c_2^*)\leq 3r^*$. The idea is to first compute $r^{DA}$ by imposing the condition that the disks should be congruent and then using $r^{DA}$ to build $(D^{DA}_1,D^{DA}_2)$. So, let $(\hat{D}_1,\hat{D}_2)$ with centers $(\hat{c}_1,\hat{c}_2)$ be a BOS having the distant assumption such that the radii of the disks are equal (disks are congruent). So, the cost of this solution would be $r^{DA}$. Here, the objective is to compute $(\hat{D}_1,\hat{D}_2)$. We first apply the algorithm of \cite{KSH} to obtain an optimal solution for the $2$-center problem on $P$ with minimum distance between their centers. In order to have this additional proximity property in the solution, we replace Hershberger's feasibility test\cite{HFT} in \cite{KSH} with Sharir's un-parallelized feasibility test in \cite{SH} (note that we can't use Sharir's algorithm in \cite{SH} because it uses simulating parallel feasibility test which makes the algorithm impractical). If the distance between the centers of this solution is equal or less than $\delta$, we set $r^{DA}$ as the cost of this solution and try to build $(D^{DA}_1,D^{DA}_2)$ (will be discussed later in the paper) based on this cost (if we couldn't build a solution with this cost satisfying the distant assumption, we would know that a BOS with this cost exists in the far distant or the nearby cases and thus, we won't miss the BOS for the problem). Otherwise, we can assume that $d(\hat{c}_1,\hat{c}_2)=\delta$ .\par
Similar to the Section 3, we build a set of constant size directed lines $\mathcal{X}$ such that for at least one directed line $\vec{x}\in\mathcal{X}$, $\hat{c}_1$ is on the negative side of $\hat{c}_2$ and the angle between $line(\hat{c}_1,\hat{c}_2)$ and $\vec{x}$ is at most $1^\circ$. Without loss of generality we can assume that $\vec{x}$ is horizontal and its positive direction is right wise. Let $v_1$ be the leftmost point of $\hat{D}_2$ not in the interior of $\hat{D}_1$ and $v_2$ be the rightmost point of $\hat{D}_1$ not in the interior of $\hat{D}_2$. If $v_1$ lies to the right of $v_2$, we already catch the BOS in the far distant case and we are done. So, assume that $v_1$ is on the left side of $v_2$. The objective here is to find a vertical line $l$ such that it separates $\hat{c}_1$ from $v_1$. First, note that the difference between the $x$-coordinates of $v_1$ and $v_2$ is at least $0.9r^*/2$ and $r^*>\Delta x/4$ where $\Delta x$ is the difference between the $x$-coordinates of the leftmost and rightmost points. Having $\Delta x$, one of the vertical lines at the distance $k\Delta x/9$ (for some $1\leq k \leq 9$) from the leftmost point will separate $\hat{c}_1$ from $v_1$. So, we can build a set $\mathcal{L}_D$ of lines such that for at least one line $l\in \mathcal{L}_D$, there exist an $x$-axis in $\mathcal{X}$ such that with respect to that, $l$ is vertical and separates $\hat{c}_1$ and $v_1$.
\begin{figure}
\begin{center}
\includegraphics[scale=0.2]{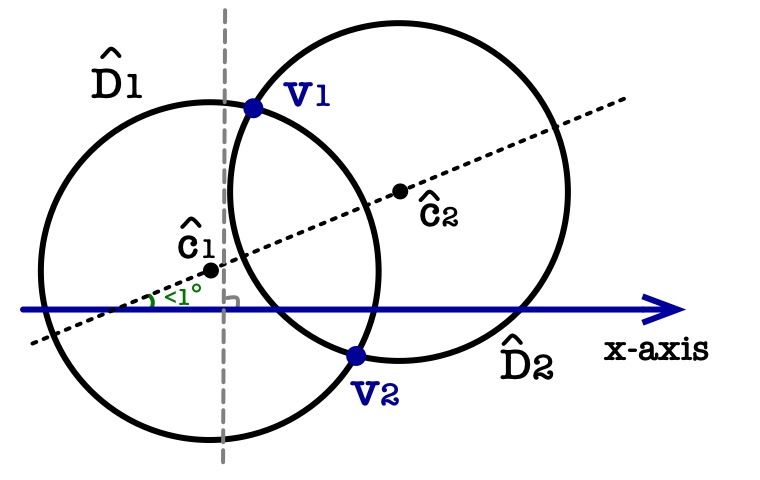}
\end{center}
\caption{Separating $\hat{c}_1$ and $v_1$ by a vertical line}
\label{fig_v1v2}
\end{figure} 

Note that all points of $P$ on the left-side of $l$ should be covered by $\hat{D}_1$ and so, we can assume that $\hat{c}_1$ is on the boundary of the intersection hull of these points at radius $r^*$. According to the assumption $dist(\hat{c}_1,\hat{c}_2)=\delta$, we can say that at radius $r^*$, the distance between the intersection hull of the points covered by $\hat{D}_1$ and the intersection hull of the points not covered by $\hat{D}_1$ is exactly $\delta$ (the distance between two intersection hulls is the minimum possible distance between their points). In the rest of this section, we discuss how we can use this property to compute $r^{DA}$ and $(\hat{D}_1,\hat{D}_2)$.\par
%-----------------------------------------------------------------
Suppose that $l$ is a given vertical line in $\mathcal{L}_D$. Denote the set of points in $P$ on the left side(resp. right side) of $l$ by $P^-$(resp. $P^+$). Also, denote the intersection hull of $P^-$ with respect to a radius $r>0$ by $H^-(r)$. Define $r_l$ as the minimum radius for which there exist a point $x\in \partial H^-(r_l)$ such that the distance between $x$ and the intersection hull of the points in $P$ not covered by $Disk(x,r_l)$ (the disk with center $x$ and radius $r_l$) is exactly $\delta$. So, in order to find $r^{DA}$ it is enough to compute $r_l$ for all $l\in \mathcal{L}_D$ and set $r^{DA}=\min\{r_l:l\in \mathcal{L}_D\}$.\par
\begin{figure}
\begin{center}
\includegraphics[scale=0.08]{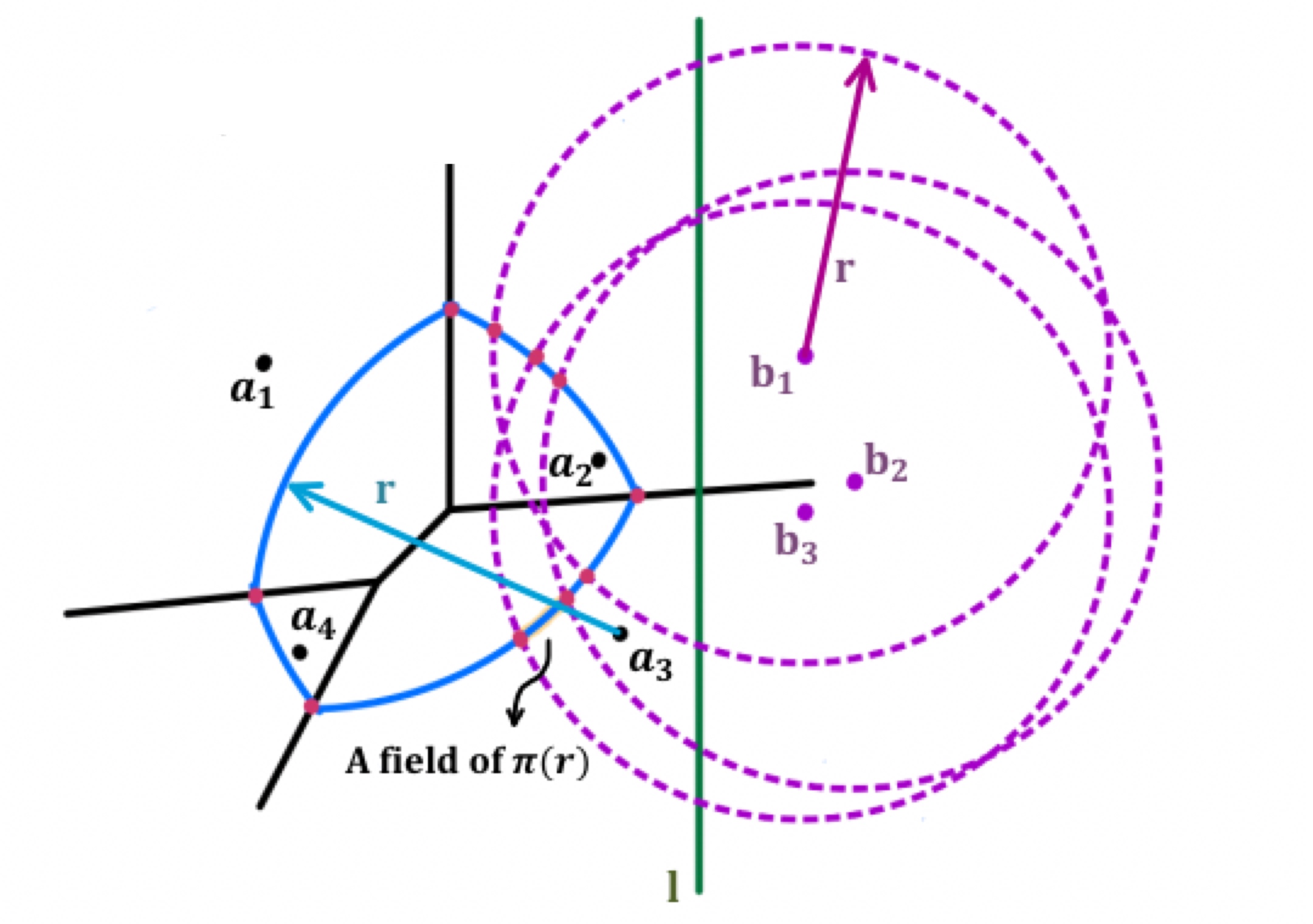}
\end{center}
\caption{The circles of $\mathcal{A}(r)$ and its induced partition $\pi(r)$ on $\partial H^-(r)$}
\label{fig_circles}
\end{figure}

In order to find $r_l$, we need a feasibility test to answer the following question: given an $r$, determine whether $r$ is greater, equal or smaller than $r_l$. Consider the set of circles $\mathcal{A}(r)=\{circle(p,r):p\in P^+\}$ ($circle(p,r)$ is the circle with center $p$ and radius $r$) and compute the intersection points of each circle of $\mathcal{A}(r)$ with $\partial H^-(r)$. These intersection points and the vertices of $\partial H^-(r)$ induce a partition on $\partial H^-(r)$. We denote this partition by $\pi(r)$ which can be considered as an alternating sequence of arc interiors and endpoints. We assume that the order is clockwise starting from its leftmost endpoint. We call each of these arc interiors and endpoints a \textit{field} of $\pi(r)$ (so, if $\pi(r)$ has $k$ arcs, it would have $2k$ fields). See Figure \ref{fig_circles}.
We observe that for each field $f$ of $\pi(r)$, $disk(x,r)$ covers a same set of points for any $x\in f$. For simplicity, we call the set of points in $P$ covered by $disk(x,r)$, the points covered by $f$ at radius $r$. So, $\pi(r)$ is a sequence of fields each covers a specific set of points. For each field $f\in \pi(r)$, denote the intersection hull of the points not covered by $f$ with respect to $r$ by $H^+_f(r)$. Note that the difference between the set of points covered by two neighbour fields in $\pi(r)$ is at most two(based on our assumption that the points are in general position). Also, we can compute the sequence of disks that enter or leave $H^+_f(r)$ when $f$ varies in $\pi(r)$ in the clockwise order. Having this sequence allows us to use the data structure of \cite{HerSur} to compute $H^+_{f'}(r)$ having $H^+_f(r)$ in $O(\log n)$ amortized time where $f'$ is a neighbour field of $f$ in $\pi(r)$. So, to do the feasibility test, we first compute $H^+_{f_0}(r)$ where $f_0$ is the first field of $\pi(r)$ and then traverse $\pi(r)$ in the clockwise order and at each field $f$, update the intersection hull of the points not covered by $f$ and compute $dist(f,H^+_f(r))$. Because both $f$ and $H^+_f(r)$ are convex, we can compute the distance between them in $O(\log n)$ time \cite{DST}. During the traversal, we stop and return \textit{greater} as soon as for a field $f$, $dist(f,H^+_f(r))<\delta$. If we reach the end of the traversal, return \textit{equal} if we've seen a field $f$ for which $dist(f,H^+_f(r))=\delta$ and we couldn't find any other field $f'$ with $dist(f',H^+_{f'}(r))<\delta$. Otherwise, we return \textit{smaller}. So, the feasibility test can be done in $O(n\log n)$ time. The Procedure RL-FTEST($r$) in Algorithm 2 represent the pseudocode of our feasibility test. 
\begin{algorithm}[H]
\caption{RL-FTEST($r$)}
\begin{algorithmic}[1]
\State Compute $H^-(r)$ and the sequence $\pi(r)=(f_0,\dots,f_m)$ on it.
\State Let $f=f_0$.
\State Compute $H^+_f(r)$ and the set of points covered by $f$ at radius $r$.
\State Let $d_{min}= dist(f,H^+_f(r))$.
\State Traverse $\pi(r)$ and at each field store the point(s) (at most two) that leave/enter the coverance.
\For {$f=f_1,\dots,f_m$}
\State Compute $H^+_f(r)$ by updating from $H^+_{f-1}(r)$.
\State $d_{min}=min\{dist(f,H^+_f(r)),d_{min}\}$. 
\If {$d_{min}<\delta$}
\State Return \textit{Greater}. 
\EndIf 
\EndFor
\If {$d_{min}=\delta$} \textbf{Return} \textit{Equal} 
\Else 
\State Return \textit{Smaller} 
\EndIf
\end{algorithmic}
\end{algorithm}
Now, we discuss our algorithm to compute $r_l$. We first build $\mathcal{F}(P^-)$ (farthest-point Voronoi diagram of $P^-$) and do a binary search on the weights (see Appendix A) of the vertices of $\mathcal{F}(P^-)$ using the above feasibility test to obtain an interval $I^*=(i_0,i_1)$ such that $r_l\in I^*$ and for each vertex $v$ of the diagram, $I^*$ does not contain the weight of $v$. Because we use $O(\log n)$ feasibility tests, the cost of obtaining $I^*$ would be $O(n\log^2 n)$. As soon as we found $I^*$, we can build $H^-(i_0)$ and the set of its arms $A^-$.\par
The idea here is simulating the propagation of $\pi(r)$ and the circles in $\mathcal{A}(r)$ when $r$ varies from $i_0$ to $i_1$. To do this, we assume that at time $t\in I^*$, the radius of $\pi$ and the circles in $\mathcal{A}$ is $t$ (if a field is an endpoint, its radius is the radius of an arc containing it). The minimum time $t$ such that $dist(f,H^+_f(t))\leq \delta$ for a field $f$ in $\pi(t)$ is actually our $r_l$. Consider $t_1,t_2\in I^*$ and let $\pi(t_1)=(f_1,\dots,f_j)$ and $\pi(t_2)=(f'_1,\dots,f'_{j'})$. We say that $\pi(t_1)$ and $\pi(t_2)$ have a same structure if $j=j'$ and for each $1\leq i\leq j$, $f_i$ and $f'_i$ cover exactly a same set of points. In this case, we consider $f_i$ and $f'_i$ as a same field in different times. Note that if $\pi(t_1)$ and $\pi(t_2)$ have different structures, for any $t\geq t_2$, $\pi(t_1)$ has a different structure from $\pi(t)$. So, we can consider a sequence of times $T=(t_0=i_0,t_1,\dots,t_k=i_1)$ for some integer $k$ such that $\pi$ has a same structure between any two consecutive times in the sequence. We call this sequence the \textit{event-sequence} and each time in this sequence an \textit{event-time}. We can see that a time $t$ is an event-time if one of the following events happens at $t$: 
\begin{enumerate}
\item
Circle in $\mathcal{A}(t)$ collides with $H^-(t)$. 
\item
Two endpoints of $\pi(t-\epsilon)$ collide at $\pi(t)$ where $\pi(t-\epsilon)$ has the structure exactly before the event for a sufficiently small $\epsilon>0$. 
\end{enumerate}
At a first type event, new fields emerge and at a second type event a field disappears (new fields may appear). Note that the number of such events is $O(n^2)$ and we can obtain $T$ by considering the intersection of each pair of circles (or disks) on $\pi$ and finally sort the times. So, we can compute $T$ (increasingly sorted) in $O(n^2\log n)$ time. Here, we are going to find an interval $T^*=(t',t'')\subseteq I^*$ such that $r_l\in T^*$ and it contains no event-time. To do this, we apply our feasibility test $O(\log n)$ times to perform a binary search on $T$ to get $T^*$. So, computing $T^*$ again costs $O(n\log^2n)$ time.\par

%Because we do not afford to store all $O(n^2)$ events, we do as follows: Initially set $T=I^*$ and for each point $p\in P^+$, we calculate the times like $t$ such that $disk(p,t)$ intersect a vertex of $\pi(t)$ or another disk $disk(p'\in P^+,t)$ on $\pi(r)$. Because the number of such events is linear, we can store and sort them to apply a binary search using the above feasibility test and finally update $T$ such that non of these event-times lie in $T$. This will cost $O(n\log ^2 n)$. We perform this binary search for all the points in $P^+$ to finally obtain $T^*$. So, the total cost of computing $T^*$ is $O(n^2\log^2 n)$.\par
Next, for each field $f\in \pi(t\in T^*)$, we store the set of points not covered by it denoted by $f^+$ and compute the farthest-point Voronoi diagram $\mathcal{F}(f^+)$ of these points. In order to find $r_l$, we need to compute $t^\delta_f$ for each field $f\in \pi(t\in T^*)$ where $t^\delta_f$ is the earliest time for which $dist(f\in \pi(t^\delta_f),H^+_f(t^\delta_f))=\delta$. Finally, we have $r_l=min\{t^\delta_f:f\in \pi(t\in T^*)\}$.\par
Here, we discuss how to compute $t^\delta_f$ for a field $f\in \pi(t\in T^*)$. First, we build $\mathcal{F}(f^+)$ in $O(n\log n)$ time. Note that having $\mathcal{F}(f^+)$, for a given $t\in T^*$, we can compute $H^+_f(t)$ with ordered arcs in linear time. Next, we apply binary search on the weights(times) of the vertices of $\mathcal{F}(f^+)$ using the algorithm of \cite{DST}(which costs $O(\log n)$ for computing the distance between two convex objects) as the feasibility test to check whether the distance between $f$ and $H^+_f$ is smaller, equal or larger than $\delta$. Let $T^*_f\subseteq T^*$ be the resulting interval. So, when $t$ varies in $T^*_f$, no arc appears or vanishes in $H^+_f(t)$. So, $T^*_f$ can be computed in $O(n\log n)$ time. Having $T^*_f$, because we have no structural change in $H^+_f$, we can compute $t^\delta_f$ in linear time (by considering the arms of $H^+_f$). So, the total cost of computing $t^\delta_f$ is $O(n\log n)$ and thus, $r_l$ can be computed in $O(n^2\log n)$ time. The procedure FIND-RL($l$) in Algorithm 3 will return $r_l$ given a line $l$. 

\begin{algorithm}[H]
\caption{FIND-RL($l$)}
\begin{algorithmic}[1]
\State Build $\mathcal{F}(P^-)$ and its weights.
\State Perform a binary search on the weights using RL-FTEST to get an interval $I^*=[i_0,i_1]$.
\State Compute the event-times sequence $T$ when $\pi(t)$ and $\mathcal{A}(t)$ expands as $t\in I^*$.
\State Perform a binary search on $T$ using RL-FTEST to get an interval $T^*$.
\State Let $f_1,\dots,f_m'$ be the fields of $\pi(t)$ when $t\in T^*$. // The fields are same as $t$ varies in $T^*$.
\For {$i$ from $1$ to $m'$}
\State Let $f^+$ be the set of points not covered by $f_i$.
\State Compute $\mathcal{F}(f^+)$ and perform a binary search on its weights with the following test:
\State Test($w$):
\State \ \ \ \ Compute $H^+_{f_i}(w)$ and let $d=dist(f_i, H^+_{f_i}(w))$.
\State \ \ \ \ \textbf{if} $d>\delta$ \textbf{then} Return \textit{Smaller}.
\State \ \ \ \ \textbf{else if} $d=\delta$ \textbf{then} Return \textit{Equal} \textbf{else} Return \textit{Greater}.
\State Let $T^*_{f_i}$ be the final interval. // When $t$ varies in $T^*_{f_i}$ the structure of $H^+_{f_i}$ doesn't change.
\State Compute the time $t^\delta_i$ for which $dist(f_i, H^+_{f_i})=\delta$.
\EndFor
\State Return $r_l$ as $min\{t^\delta_i\;:\;1\leq i\leq m'\}$.
\end{algorithmic} 
\end{algorithm}

Having $r^{DA}$, we can obtain a BOS for it as follows: for each $l\in \mathcal{L}_2$, we consider two assumptions: first, the determining disk is on the left side of $l$ and second it is on the right side of $l$. For the first case, we consider $H^-(r^{DA})$ and circles in $\mathcal{A}$ as fixed objects (not propagating) and for each field $f$, we propagate $H^+_f$ up to radius $R^{DA}$ based on the method we explained above and compute the minimum radius(time) for which $dist(H^-(r^{DA}),H^+_f)=\delta$. Let $r'_l$ be the minimum such radius and infinite if it doesn't exist. For the second case, we propagate both circles in $\mathcal{A}$ and fields of $H^-(i_0)$ to obtain event-times but we build all $H^+_f$s according to the fixed radius $r^{DA}$ and follow the above algorithm to obtain a radius $r''_l$. Comparing $r'_l$, $r''_l$ and their solutions, we obtain a BOS assuming $l$ is a correct line. Comparing the results for all $l\in \mathcal{L}_2$ will give us a BOS $(D^{DA}_1,D^{DA}_2)$ having the distant assumption with total complexity of $O(n^2\log n)$.\newpage

\small
\bibliographystyle{abbrv} 

\begin{thebibliography}{9}
\bibitem{AP}
Agarwal PK, Procopiuc CM. Exact and approximation algorithms for clustering. \textit{Algorithmica.} 2002 Jun 1;33(2):201-26
\bibitem{ASH}
Agarwal PK, Sharir M. Planar geometric location problems. \textit{Algorithmica.} 1994 Feb;11(2):185-95.
\bibitem{ASH2}
Agarwal PK, Sharir M. Efficient algorithms for geometric optimization. \textit{ACM Computing Surveys (CSUR)}. 1998 Dec 1;30(4):412-58.
\bibitem{AROR}
Arora S. Nearly linear time approximation schemes for Euclidean TSP and other geometric problems. In \textit{Proceedings 38th Annual Symposium on Foundations of Computer Science} 1997 Oct 20 (pp. 554-563). IEEE.
\bibitem{CHN}
Chan TM. More planar two-center algorithms. \textit{Computational Geometry.} 1999 Sep 1;13(3):189-98.
\bibitem{DST}
Chin F. Optimal algorithms for the intersection and the minimum distance problems between planar polygons. \textit{IEEE Transactions on Computers.} 1983 Dec 1(12):1203-7.
\bibitem{CHOO}
Cho K, Oh E. Optimal algorithm for the planar two-center problem. arXiv preprint arXiv:2007.08784. 2020 Jul 17
\bibitem{CAHN}
Choi J, Ahn HK. Efficient planar two-center algorithms. \textit{Computational Geometry.} 2021 Apr 2:101768.
\bibitem{CL}
Cole R. Slowing down sorting networks to obtain faster sorting algorithms. \textit{Journal of the ACM (JACM)}. 1987 Jan 1;34(1):200-8.
\bibitem{DRENZ}
Drezner Z. The planar two-center and two-median problems. \textit{Transportation Science.} 1984 Nov;18(4):351-61.
\bibitem{EPS}
Eppstein D. Faster construction of planar two-centers. In \textit{Proc. of the 8th Annual ACM-SIAM Symposium on Discrete Algorithms (SODA)}, pages 131–138, 1997.
\bibitem{GUD}
Gudmundsson J, Haverkort H, Park SM, Shin CS, Wolff A. Facility location and the geometric minimum-diameter spanning tree. \textit{Computational Geometry}. 2004 Jan 1;27(1):87-106.
\bibitem{HFT}
Hershberger J. A faster algorithm for the two-center decision problem. \textit{Information processing letters.} 1993 Aug 9;47(1):23-9.
\bibitem{HER2}
Hershberger J, Suri S. Efficient computation of Euclidean shortest paths in the plane. In \textit{Proceedings of 1993 IEEE 34th Annual Foundations of Computer Science} 1993 Nov 3 (pp. 508-517). IEEE.
\bibitem{HerSur}
Hershberger J, Suri S. Off-line maintenance of planar configurations. \textit{Journal of algorithms.} 1996 Nov 1;21(3):453-75.
\bibitem{ACN0}
Huang CH. Some problems on radius-weighted model of packet radio network (Doctoral dissertation, Ph. D. Dissertation, Dept. of Comput. Sci., Tsing Hua Univ., Hsinchu, Taiwan).
\bibitem{ACN2}
Huang PH, Tsai YT, Tang CY. A near-quadratic algorithm for the alpha-connected two-center problem. \textit{Journal of information science and engineering.} 2006 Nov 1;22(6):1317.
\bibitem{ACN1}
Huang PH, Te Tsai Y, Tang CY. A fast algorithm for the alpha-connected two-center decision problem. \textit{Information Processing Letters.} 2003 Feb 28;85(4):205-10.
\bibitem{HW}
Hwang RZ, Lee RC, Chang RC. The slab dividing approach to solve the Euclidean P-Center problem. \textit{Algorithmica}. 1993 Jan 1;9(1):1-22.
\textit{SIAM Journal on Applied Mathematics.} 1979 Dec;37(3):513-38.
\bibitem{KSH}
Katz MJ, Sharir M. An expander-based approach to geometric optimization. \textit{SIAM Journal on Computing.} 1997 Oct;26(5):1384-408.
\bibitem{CAP1}
Khuller S, Sussmann YJ. The capacitated k-center problem. \textit{SIAM Journal on Discrete Mathematics.} 2000;13(3):403-18.
\bibitem{CAP2}
Lim A, Rodrigues B, Wang F, Xu Z. k-Center problems with minimum coverage. \textit{Theoretical Computer Science.} 2005 Feb 28;332(1-3):1-7.
\bibitem{DNK}
Gowda I, Kirkpatrick D, Lee D, Naamad A. Dynamic voronoi diagrams. \textit{IEEE Transactions on Information Theory.} 1983 Sep;29(5):724-31.
\bibitem{MG}
Megiddo N. Linear-time algorithms for linear programming in $\mathbb{R}^3$ and related problems. \textit{SIAM journal on computing.} 1983 Nov;12(4):759-76.
\bibitem{MGPS}
Megiddo N. Applying parallel computation algorithms in the design of serial algorithms. \textit{Journal of the ACM (JACM)}. 1983 Oct 1;30(4):852-65.
\bibitem{OVER}
Overmars MH, Van Leeuwen J. Maintenance of configurations in the plane. \textit{Journal of computer and System Sciences.} 1981 Oct 1;23(2):166-204.
\bibitem{SH}
Sharir M. A near-linear algorithm for the planar 2-center problem. \textit{Discrete and Computational Geometry.} 1997 Sep 1;18(2):125-34.
\bibitem{HAND}
Toth CD, O'Rourke J, Goodman JE, editors. Handbook of discrete and computational geometry. CRC press; 2017 Nov 22.
\bibitem{HT}
Wang H. On the Planar Two-Center Problem and intersection Hulls. arXiv preprint arXiv:2002.07945. 2020 Feb 19.
\end{thebibliography}
\newpage
\section*{Appendix A: A review on the farthest-point Voronoi Diagram of a set of points and their properties.}
\begin{figure}
\begin{center}
\includegraphics[scale=0.4]{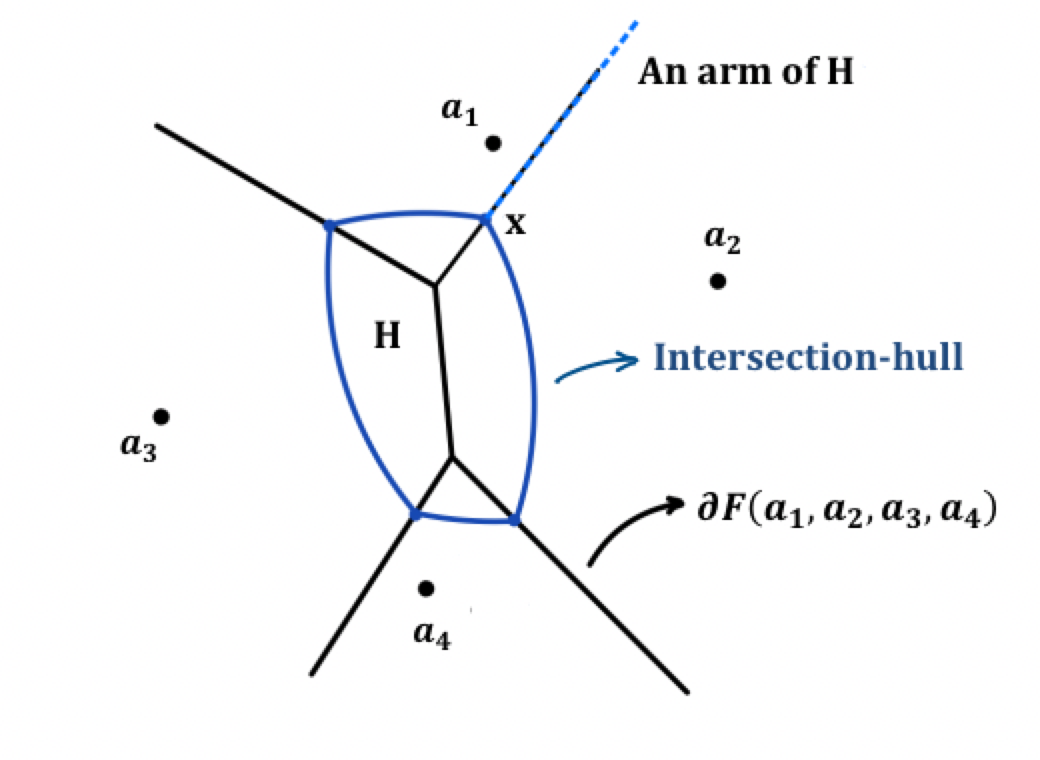}
\end{center}
\caption{The farthest-point Voronoi diagram of a set of points and its intersection hull at some radius $r$.}
\label{Fig1}
\end{figure}

For a given set of points $A=\{a_i:1\leq i\leq m\}$, the farthest-point Voronoi diagram of $A$ denoted by $\mathcal{F}(A)$ is the partition of the plane into a set of disjoint-interior \textit{cells} $\{C(a_i):a_i\in A\}$ such that $C(a_i)$ is the set of points in the plane for which no point of $A$ is farther from them than $a_i$. We say $a_i$ is the farthest point of $C(a_i)$ and call it the \textit{site} of the cell $C(a_i)$. For each point $x\in C(a_i)$, the \textit{weight} of $x$ is defined as its distance to $a_i$ and we denote it by $w(x)$. Note that $disk(x,w(x))$ (the disk with center $x$ and radius $w(x)$) covers all the points of $A$. It is easy to see that $\partial \mathcal{F}(A)$ (the set of boundaries between cells of $\mathcal{F}(A)$) consists of a set of line segments or half-lines, or it is just a line which we call them the \textit{edges} of $\partial \mathcal{F}(A)$. For each $e\in \partial \mathcal{F}(A)$, there exists a unique pair $(a_i,a_j)$ of points of $A$ such that for any point $x$ on the interior of $e$, we have $dist(x,a_i)=dist(x,a_j)$ and no point of $A$ is farther than these points from $x$. We call $a_i$ and $a_j$ the \textit{generators} of $e$. Note that $e$ lies on the perpendicular bisector of the  $seg(a_i,a_j)$ (the line segment connecting $a_i$ and $a_j$). If $v$ is an endpoint of $e$ (in this case we call $v$ a vertex), $v$ has three points in $A$ all farthest from $v$ (and no more because of our assumption that no four points are on a circle). We also call these points the generators of $v$. Given a start and an end point on $\partial \mathcal{F}(A)$, its corresponding path on $\partial \mathcal{F}(A)$ is the portion of $\partial \mathcal{F}(A)$ between two points directed from the start point toward the end point. Note that such a path is unique otherwise a cell of the $\mathcal{F}(A)$ would be bounded which is not possible \cite{HAND}. 
\begin{obs}
The center of the minimum enclosing disk of $A$ is the minimum weight point on $\partial \mathcal{F}(A)$.
\label{obs_center_minimum_weight}
\end{obs}
We call the center of the minimum enclosing disk of $A$ the \textit{root} of $\partial \mathcal{F}(A)$ and it is unique.
\begin{prop}
Let $p=[r,b]$ be a path on $\partial \mathcal{F}(A)$ where $r$ is its root. Then, the weight of the points on $p$ change monotonically increasing from $r$ to $b$.
\label{prop_weight_change}
\end{prop}
\textbf{Proof.}
First, note that the root is unique. So, for each edge $e$ in the path with generators $a_i$ and $a_j$, the midpoint of $seg(a_i,a_j)$ can not lie on the interior of $e$. On the other hand, $e$ is a subset of the perpendicular bisector of $seg(a_i,a_j)$ and the weight of each point on $e$ is its distance to $a_i$ (which is the same as its distance to $a_j$). So, the weight on $e$ should change monotonically as we move from one of its endpoints to another. Now, suppose that the proposition is not true. Then, there must be a first edge $ht$ (direction is along $p$) on the path such that as we move from $h$ to $t$, the weight decreases. This means that the vertex $h$ should have local maximum weight on the path which is contradiction because if we slightly move from $h$, the distance with one of its generators should be increased. $\hfill \square$\\\\
The intersection hull of $A$ at radius $r$ is defined as $\cap_{a\in A}disk(a,r)$. Lets denote the intersection hull of $A$ at radius $r$ by $H_A(r)$. We can easily see that $H_A(r)$ is composed of a set of circle arcs with radius $r$ with end points at the edges of $\partial \mathcal{F}(A)$. So, if we start from the leftmost endpoint of $H(r)$ and traverse its arcs clockwise, we obtain a unique sequence of arcs. We refer to this sequence as the \textit{arc-sequence} of $H_A(r)$ and denote it by $Seq(H(r))$. Let $x$ be a vertex of $H_A(r)$. Suppose that $x$ lies on an edge $e$ of $\mathcal{F}(A)$. We call the half-line from $x$ along $e$ that does not intersect the interior of $H_A(r)$ the \textit{arm} of $H_A(r)$ from $x$. See Firgure \ref{Fig1}.\newpage
 \newpage
\newpage
\section*{Appendix B: Proofs}
\begin{figure}[H]
\begin{center}
\includegraphics[scale=0.1]{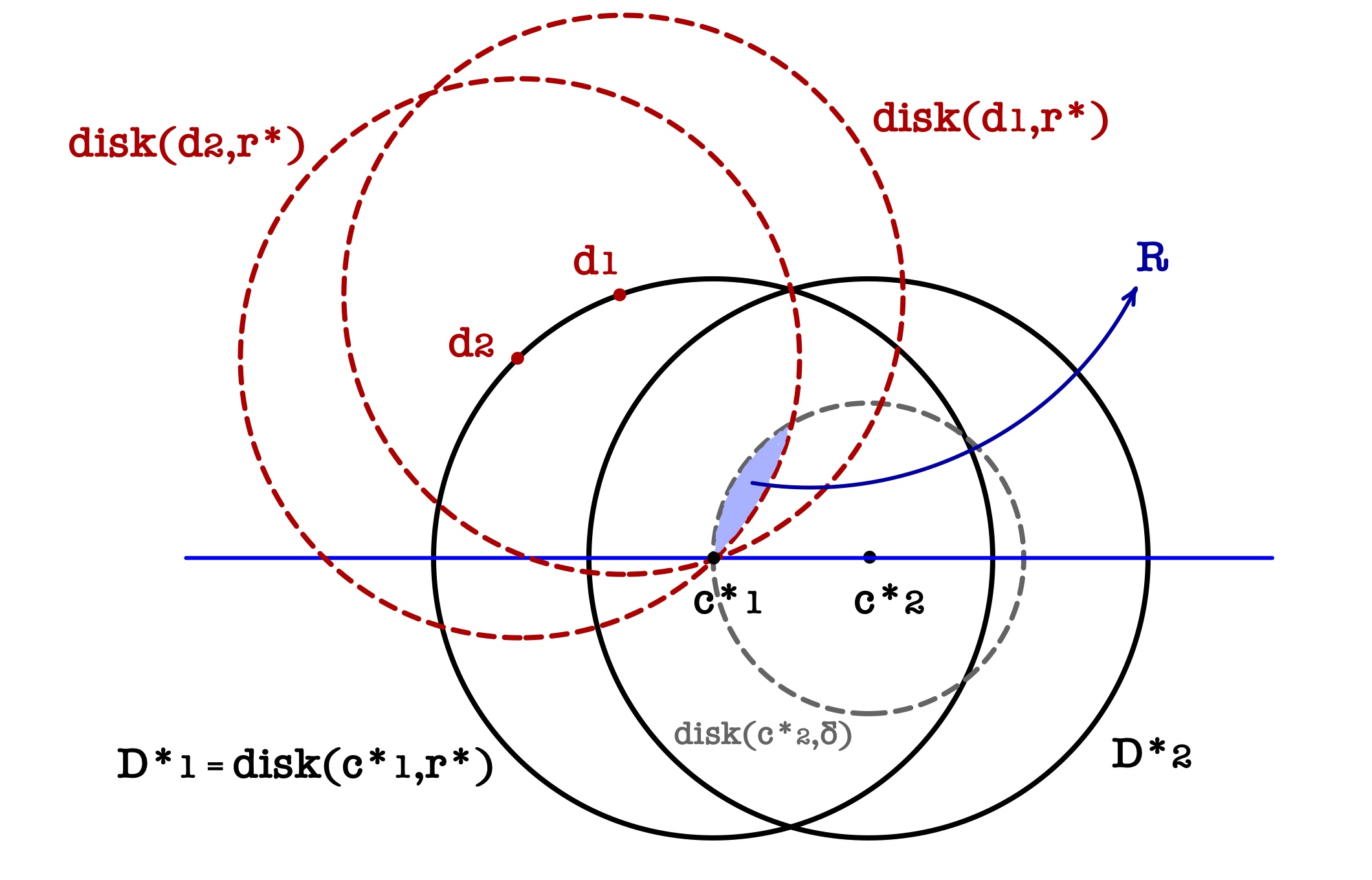}
\end{center}
\caption{If $d_1$ and $d_2$ are on a same side of $line(c^{i,j}_+, c^{i,j}_-)$, $(D^{i,j}_-,D^{i,j}_+)$ can't be best optimal}
\label{fig_dom_dif_sides}
\end{figure}

\textbf{Proof of Proposition \ref{prop_dif_int}}. First, we prove that $D^*_1$ should have a pair of dominating points each on the different sides of $line(c^*_1,c^*_2)$. Note that if $D^*_1$ has three dominating points on one side of $line(c^*_1,c^*_2)$, their induced triangle can't cover $c^*_1$ and based on our general assumption such situation can't happen. Now, suppose that $D^*_1$ has two dominating points $d_1$ and $d_2$ both on a same side of $line(c^*_1,c^*_2)$. Because $D^*_1$ is not MED (it has only two dominating points), the distance between $c^*_1$ and $c^*_2$ should be exactly $\delta$ (otherwise, move $c^*_1$ toward the dominating points to get better solution). Because $d_1$ and $d_2$ are on a same side of $line(c^*_1,c^*_2)$, the region $R:=disk(d_1,r^*)\cap disk(d_2,r^*)\cap disk(c^*_2,\delta)$ is not empty. So, if we slightly move $c^*_1$ into  $R$, we would get a better solution which is contradiction (See Figure \ref{fig_dom_dif_sides}). 
For the second statement, again if $D^*_1$ has three dominating points such that no pair of them intersect $seg(c^*_1,c^*_2)$, their induced triangle can not contain $c^*_1$ which is contradiction. Now, suppose that $D^*_1$ has two dominating points $d_1$ and $d_2$ such that their connecting segment intersect $seg(c^*_1,c^*_2)$. Because $seg(d_1,d_2)$ has non-empty intersection with the interior of $disk(c^*_2,\delta)$ and this disk is tangent to $c^*_1$, we can slightly move $c^*_1$ toward the mid-point of $seg(d_1,d_2)$ while we are still inside $disk(c^*_2,\delta)$ to get a better solution.  $\hfill\square$ \\\\
\noindent \textbf{Proof of Proposition \ref{prop_123_matrix}}. 1,2) We prove the first statement and the second statement is similar. Because $M^+[i,j]$ is non-critical, its dominating points make a triangle for $c^{i,j}_+$ (their induced triangle covers $c^{i,j}_+$). On the other hand, $P^{i,j}_+\subset P^{i',j'}_+$. So, $D^{i',j'}_+$ contains the dominating points of $D^{i,j}_+$ and its radius should be greater or equal than $D^{i,j}_+$ which gives Proposition. 3) Suppose $M^+[i,j]$ is critical and $M^+[i,j]>radius(D^{i,j}_-)$. Then, we can slightly move both centers toward the dominating points of $D^{i,j}_+$ and get a solution with reduced cost which contradicts best optimality of $(D^{i,j}_+,D^{i,j}_-)$ 4) If $M^+[i,j]> radius(D^{i,j}_-) $, based on case 3, $M^+[i,j]$ should be non-critical while we assumed it is critical. 5) If $d(c^{i,j}_-,c^{i,j}_+)<\delta$, we can slightly move $c^{i,j}_+$ toward the dominating points of $D^{i,j}_+$ without violating the PCC and reduce the radius of $D^{i,j}_+$ which again contradicts best optimality.  $\hfill \square$\\\\
\newpage
\section*{Appendix C: Proof of the Theorem \ref{main_theorem} }
\begin{figure}[H]
\begin{center}
\includegraphics[scale=0.1]{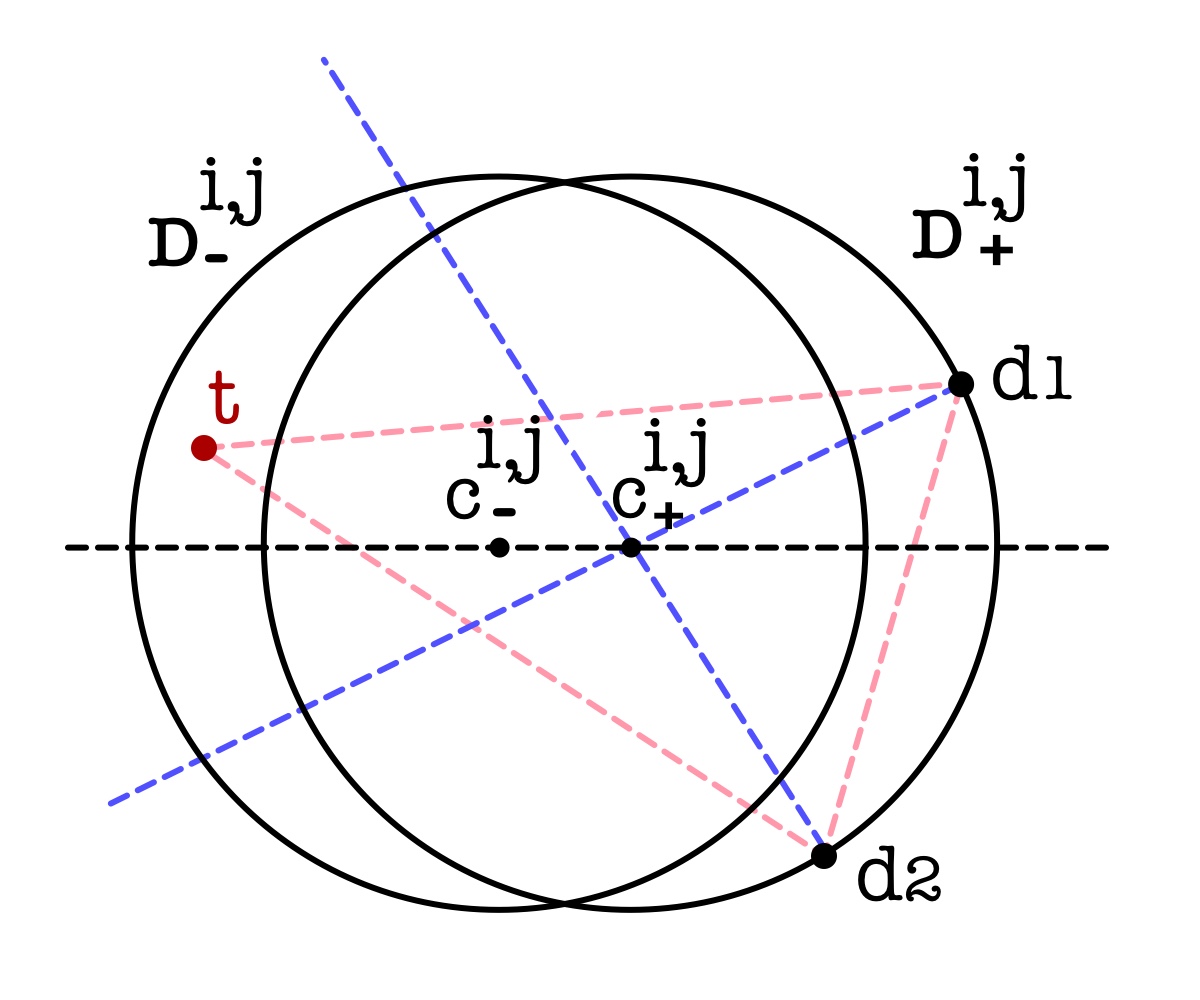}
\end{center}
\caption{$d_1,d_2$ and $x$ make a triangle for $c^{i,j}_+$}
\label{fig_triangle}
\end{figure}

Lets $x$ and $y$ be two points in the plane. We recall that the directed line passing from $x$ and $y$ directed from $x$ to $y$ is denoted by $line(x,y)$. Also, we denote the half-line from $x$ passing $y$ by $half\text{-}line(x,y)$. In this section, when we say first,  second, third and fourth quarter of a point $t$ with respect to some directed line $l$ we mean the first, second, third and fourth quarter of the plane when we consider $t$ as the origin and the directed line parallel to $l$ passing $t$ as the $x$-axis. We prove the theorem by providing several propositions. For simplicity, when we assign a letter to a geometric object inside a proof, the scope of that notation is only inside that proof and we may assign that letter to another object later. In addition, when we state a proposition or observation, we mean if the proposition or observation is false, the statement is either impossible or the theorem follows. So, we can assume that after a proposition or observation, its statement is always true. Also, when we say one disk is smaller than another disk, we mean smaller or equal. Let $c$ be the center of some disk $D$. We say that a set of three points $T$ make a triangle for $c$ if their induced triangle covers $c$ and any disk that covers $T$ has a radius greater than $radius(D)$. 
%\begin{prop}
%Suppose that $M^+[i,j]$ and $M^-[i,j]$ are both critical (and so valid) and $D^{i,j}_+$ (resp. $D^{i,j}_-$) has two dominating points $d_1$ and $d_2$. Then $d_1$ and $d_2$ should be on different sides of $line(c^{i,j}_+, c^{i,j}_-)$.
%\label{prop_dom_dif_sides}
%\end{prop}
%\textbf{Proof.} We proceed by contradiction. Suppose that both $d_1$ and $d_2$ are on a same side of $line(c^{i,j}_+, c^{i,j}_-)$. First, note that because $D^{i,j}_+$ is not the MED, the distance between $c^{i,j}_+$ and $c^{i,j}_-$ should be exactly $\delta$ and based on Proposition \ref{prop_123_matrix}, $M^+[i,j]=M^-[i,j]$ (the radii of the disks are equal). Now, because both $d_1$ and $d_2$ are on a same side of $line(c^{i,j}_+, c^{i,j}_-)$, $disk(d_1,M^+[i,j])\cap disk(d_2,M^+[i,j])\cap disk(c^{i,j}_-,\delta)$ has non-empty interior around $c^{i,j}_+$ (see Figure \ref{fig_dom_dif_sides}). This means that if we put $c^{i,j}_+$ at this interior, its distance to both $d_1$ and $d_2$ will be decreased while it still satisfies the PCC which contradicts best optimallity of $(D^{i,j}_-,D^{i,j}_+).\hfill\square$
%\begin{figure}[H]
%\begin{center}
%\includegraphics[scale=0.35]{FIG_PROP7.PNG}
%\end{center}
%\caption{If $d_1$ and $d_2$ are on a same side of $line(c^{i,j}_+, c^{i,j}_-)$, $(D^{i,j}_-,D^{i,j}_+)$ can't be best optimal}
%\label{fig_dom_dif_sides}
%\end{figure} 
\begin{obs}
Let $d_1$ and $d_2$ be the dominating points of $D^{i,j}_+$. Then for any point $t\notin D^{i,j}_+$ inside the cone induced by two half-lines from $c^{i,j}_+$ along $line(d_1,c^{i,j}_+)$ and $line(d_2,c^{i,j}_+)$ containing $c^{i,j}_-$, the points $d_1d_2t$ makes a triangle for $c^{i,j}_+$.
\label{obs_triangle}
\end{obs}
A similar statement is also true for $D^{i,j}_-$ and their dominating points. Figure \ref{fig_triangle} shows an example for Observation \ref{obs_triangle}.
Let $M^+[\hat{i},\bar{j}]$ be a given doubly-marked element. Also, let $M^+[\bar{i},\bar{j}]$ and $M^+[\hat{i},\hat{j}]$ be the elements for which we marked $M^+[\hat{i},\bar{j}]$ when we evaluated them in the top-right and bottom-left initial searches respectively. For simplicity of notation, henceforth we denote $D^{\hat{i},\bar{j}}_+$, $D^{\hat{i},\bar{j}}_-$, $c^{\hat{i},\bar{j}}_+$ and $c^{\hat{i},\bar{j}}_-$ by $D'_+$, $D'_-$, $c'_+$ and $c'_-$ respectively. Similarly, we denote $D^{\bar{i},\bar{j}}_+$, $D^{\bar{i},\bar{j}}_-$, $c^{\bar{i},\bar{j}}_+$, $c^{\bar{i},\bar{j}}_-$ by $\bar{D}_+$, $\bar{D}_-$, $\bar{c}_+$, $\bar{c}_-$ and $D^{\hat{i},\hat{j}}_+$, $D^{\hat{i},\hat{j}}_-$, $c^{\hat{i},\hat{j}}_+$, $c^{\hat{i},\hat{j}}_-$ by $\hat{D}_+$, $\hat{D}_-$, $\hat{c}_+$, $\hat{c}_-$ respectively. Note that based on our assumptions all of these disks has exactly two dominating points.\par
We denote the dominating points of $\bar{D}_+$ by $a$ and $b$, $\bar{D}_-$ by $x$ and $y$, $\hat{D}_+$ by $c$ and $d$, $\hat{D}_-$ by $w$ and $u$, $D'_+$ by $h_1$ and $h_2$ and finally $D'_-$ by $h'_1$ and $h'_2$. Let $a$ and $c$ be the two dominating points of $\bar{D}_+$ and $\hat{D}_+$ who lie on the opposite side of $x$ and $w$ with respect to $line(\bar{c}_-,\bar{c}_+)$ and $line(\hat{c}_-,\hat{c}_+)$ respectively. Note that $h'_1$ and $h'_2$ should be in both $P^{\bar{i},\bar{j}}_-$ and $P^{\hat{i},\hat{j}}_-$ (because if they are in the positive side, they can't be dominating points of $D'_-$). $a,b,c$ and $d$ should be in $P^{\hat{i},\bar{j}}_+$ and so covered by $D'_+$ (because we only add points 
to the positive side when we walk on $M^+$ from left to right or top to bottom). Also, suppose that $x$ and $w$ are the dominating points of $\bar{D}_-$ and $\hat{D}_-$ respectively who are moved to the positive side in the $(\hat{i},\bar{j})$-partition. So, we can assume that $y$ and $u$ are not in $P^{\hat{i},\bar{j}}_+$. This is because if for example $y\in P^{\hat{i},\bar{j}}_+$, $D'_+$ should cover $a,b,x$ and $y$ which are the all dominating points in the pair $(\bar{D}_-,\bar{D}_+)$. This means that the radius of $D'_+$ and any positive disk of $(i',j')$-partition with $i'\geq \hat{i}$ and $j'\geq \bar{j}$ is greater than the radius of $\bar{D}_-$ and so, we can discard them and the theorem follows (cases 2 and 3 in the theorem). Note that $x$ is $p$-type and $w$ is $q$-type (because $x$(resp. $w$) is moved to the positive side when we walk on a column(resp. row) of $M^+$) also $y,u$ should be covered by $D'_-$. Furthermore, $x\in P^{\hat{i},\hat{j}}_+$ because $x$ is $p$-type and if $x\in P^{\hat{i},\hat{j}}_-$, we can not bring it into the positive side in the $(\hat{i},\bar{j})$-partition by walking on the $\hat{i}^{th}$-row. Similarly, $w\in P^{\bar{i},\bar{j}}_+$. We assume that $u$ is covered by $\bar{D}_-$ because if $u\in P^{\bar{i},\bar{j}}_+$, then $u$ should also be in $P^{\hat{i},\bar{j}}_+$ which means $D'_+$ would cover $c,d,w,u$ and the theorem follows. Similarly, $y$ should be covered by $\hat{D}_-$. Note that the intersection of the disks are non-empty because we are in the nearby case. So, we can consider a point inside the intersection of the disks and have an angular clockwise and counter-clockwise order for all the dominating and intersection points of the disks.
%See Figure \ref{fig_big_example} as an example of such configuration of points. The blue disks are $\bar{D}_+$ and $\bar{D}_-$, the green disks are $\hat{D}_+$ and $\hat{D}_-$ and the red disks are $D'_-$ and $D'_+$.
%\begin{figure}[H]
%\begin{center}
%\includegraphics[scale=0.5]{FIG_BIG_EXAMPLE.PNG}
%\end{center}
%\caption{Configuration of $\bar{D}_-,\bar{D}_+,\hat{D}_-,\hat{D}_+,D'_-,D'_+$ and their dominating points in a doubly-marked element.}
%\label{fig_big_example}
%\end{figure} 
We consider two cases for $M^+[\hat{i},\bar{j}]$ and prove the theorem for each case separately. 
\subsection*{Case 1: $M^+[\hat{i},\bar{j}]> max\{M^+[\bar{i},\bar{j}],M^+[\hat{i},\hat{j}]\}$} 
\begin{figure}[h]
\begin{center}
\includegraphics[scale=0.12]{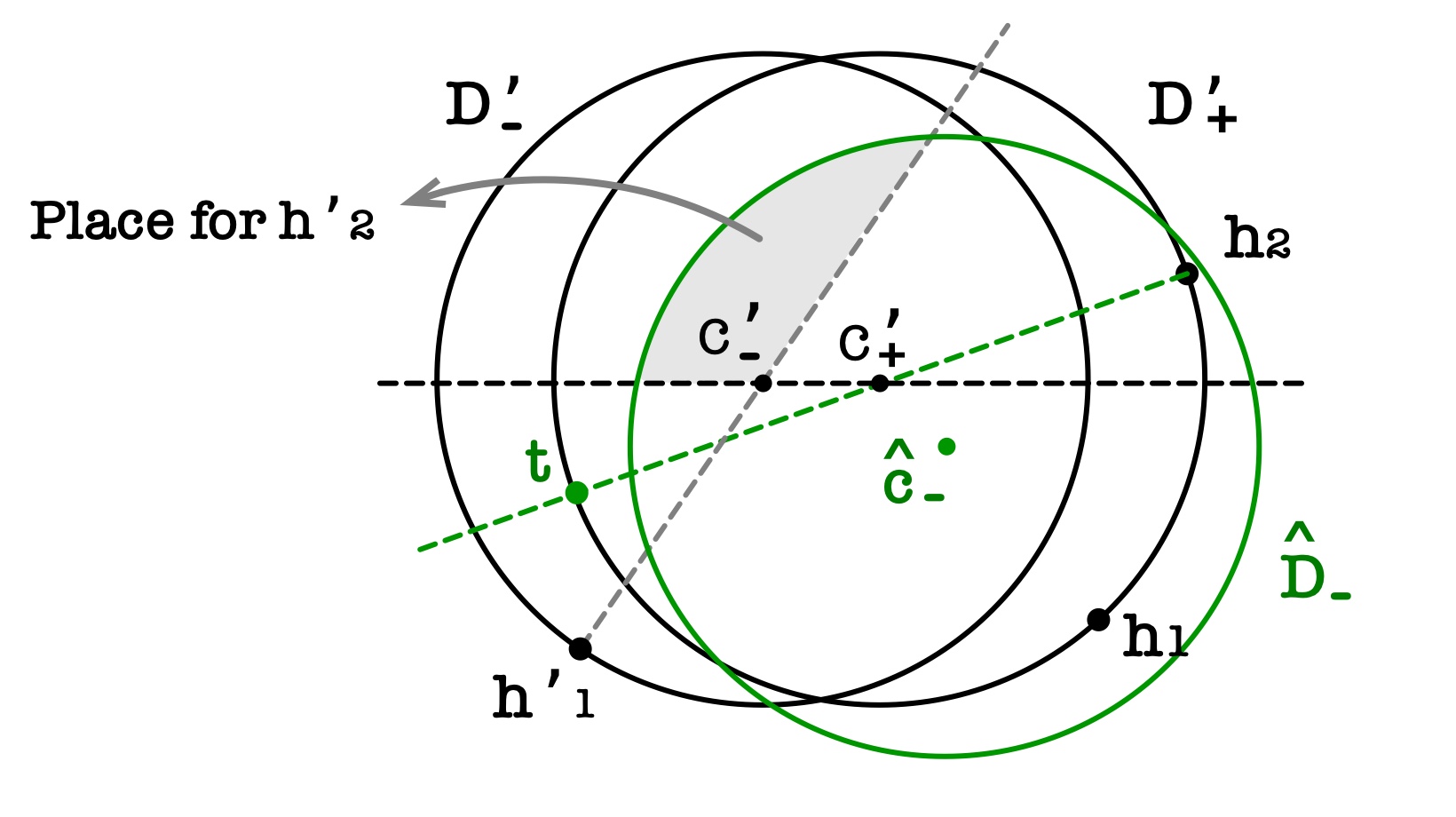}
\end{center}
\caption{Proof of Sub-case 2 in Case 1.}
\label{fig_case1}
\end{figure}

According to Proposition \ref{prop_dom_dif_sides} part 2, both $h'_1$ and $h'_2$ can't be on $D'_+$. Let $h'_1$ be the one outside $D'_+$ (the case $h'_2$ is outside $D'_+$ is similar). We show that $h'_1h_1h_2$ makes a triangle for $c'_+$. Suppose not. Based on Observation \ref{obs_triangle}, for one of $h_1$ or $h_2$ namely $h_2$, $h'_1$ should be on the opposite side of $h_2$ with respect to $line(c'_-,c'_+)$ (see Figure \ref{fig_case1}). Also, $h'_1$ and $c'_-$ should lie on opposite sides of $line(h_2,c'_+)$. Note that $h_2$ should lie outside of $D'_-$ otherwise, we can't place $h'_1$ having these conditions. \par
Without loss of generality, we assume that $h_1$ is $p$-type (the case $h_1$ is $q$-type is similar). Because $h_1$ is $p$-type, $h_1\in \hat{D}_+$ (if $h_1$ was $q$-type, we would have $h_1\in \bar{D}_+$). This is because when we traverse on a row, we only move $q$-type points to the positive side. On the other hand, both $h'_1$ and $h'_2$ should be covered by $\hat{D}_-$. Based on this situation, we have two sub-cases:\\\\
\textbf{sub-case 1: $h_2\in \hat{D}_+$.} In this sub-case, because $\{h'_1,h'_2\} \in \hat{D}_-$ and $\{h_1,h_2\}\in \hat{D}_+$, $M^+[\hat{i},\bar{j}]$ can't be greater than $M^+[\hat{i},\hat{j}]$ which is contradiction.\\\\
\textbf{sub-case 2: $h_2\in \hat{D}_-$.} Let $t$ be the intersection point of $half\text{-}line(h_2,c'_+)$ and $D'_+$ (see Figure \ref{fig_case1}). Note that $\hat{D}_-$ covers $h'_1$ and it can not cover $t$ (because it is smaller). So, $\hat{c}_-$ should lie on the side of $line(h_2,c'_+)$ that has $h'_1$. Also, $\hat{D}_-$ covers all points of $P^{\hat{i},\bar{j}}_-$. % We prove that $c'_-$ should also be on the same side of $line(h_2,c'_+)$ that has $h'_1$ which is contradiction. Suppose not. 
Now, $h'_1$ should be on the third quarter of $c'_+$ with respect to $line(h_2,c'_+)$. This is because $h'_1$ is outside $D'_+$ (the way we chose $h'_1$) and $h_2$ is on the opposite side of $line(c'_-,c'_+)$ with respect to $h'_1$. 
%We know that $c'_-$ should be on $Circle(h'_1,M^+[\hat{i},\bar{j}])$ and this circle does not intersect the first quarter of $c'_+$ with respect to $line(c'_+,h_2)$.
Also, $h'_1$ and $h'_2$ are on opposite sides of $line(h_2,c'_+)$ (because of Proposition \ref{prop_dom_dif_sides}). Now, for any point $z$ inside $\hat{D}_-$, if $h'_1z$ intersects the 
$half\text{-}line(c'_-,c'_+)$, it can't be $h'_2$ (again Proposition \ref{prop_dom_dif_sides} second part) and if it doesn't, $|zc'_-|<|h'_1c'_-|=M^+[\hat{i},\bar{j}]$ which again means that $z$ can't be $h'_2$. So, we don't have any place for $h'_2$ which is contradiction (see Figure \ref{fig_case1}).
\subsection*{Case 2: $M^+[\hat{i},\bar{j}]\leq max\{M^+[\bar{i},\bar{j}],M^+[\hat{i},\hat{j}]\}$} 
In this section, we assume that $M^+[\hat{i},\bar{j}]\leq M^+[\bar{i},\bar{j}]$ and the case $M^+[\hat{i},\bar{j}]\leq M^+[\hat{i},\hat{j}]$ is similar. Henceforth, we consider $line(\bar{c}_-,\bar{c}_+)$ as the $x$-axis unless we say otherwise. We consider two sub-cases based on the position of $x$ with respect to $\bar{D}_+$. For simplicity, when we provide a proposition within each case or sub-case, we include the assumptions of the case or sub-case in the proposition. 
\subsubsection{\underline{Sub-case 1: $\mathbf{x\notin \bar{D}_+}$.}\\\\}
\begin{figure}[h]
\begin{center}
\includegraphics[scale=0.12]{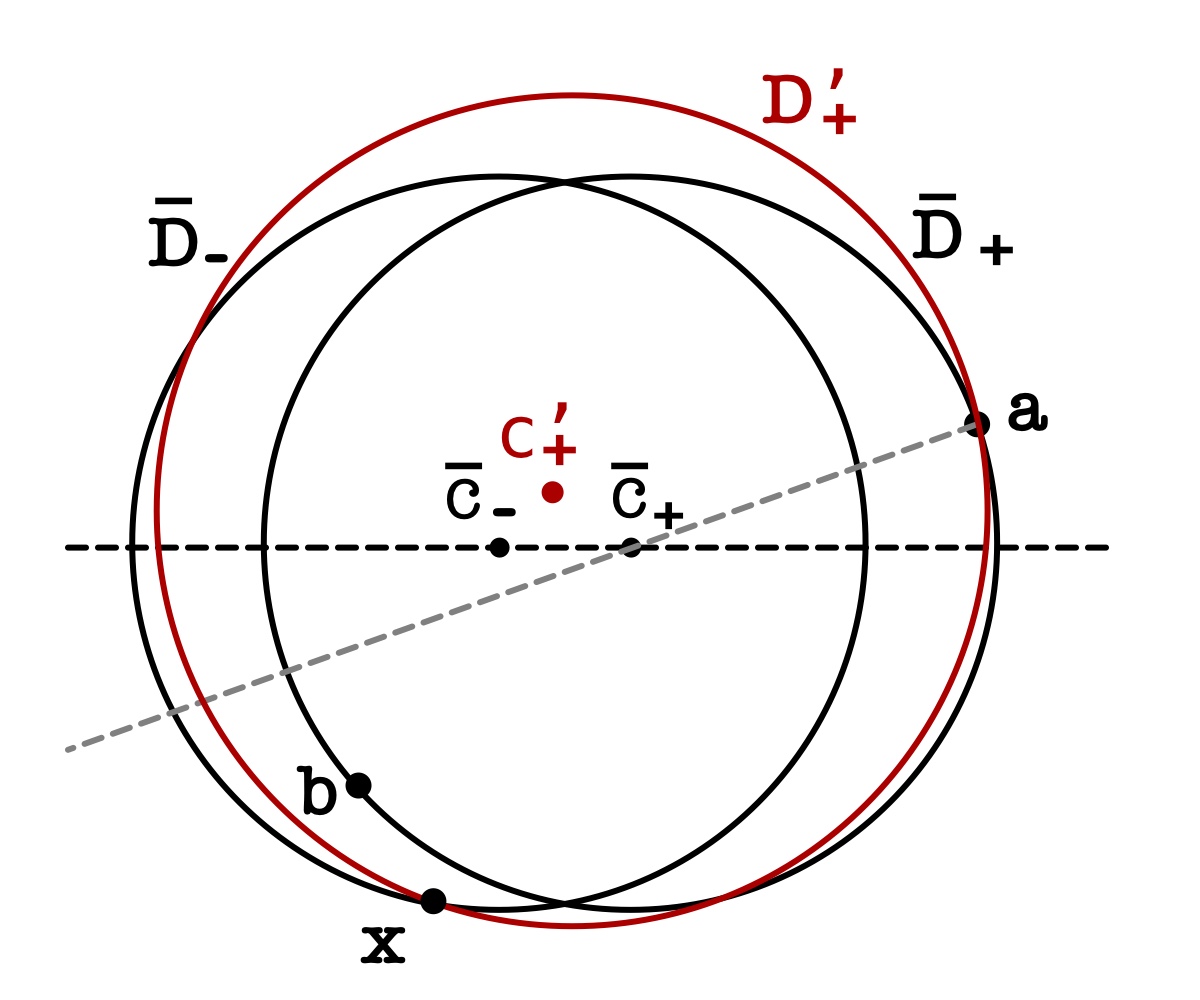}
\end{center}
\caption{Proof of Proposition \ref{prop_lower_right}.}
\label{fig_prop8}
\end{figure}

\begin{figure}[H]
\begin{center}
\includegraphics[scale=0.12]{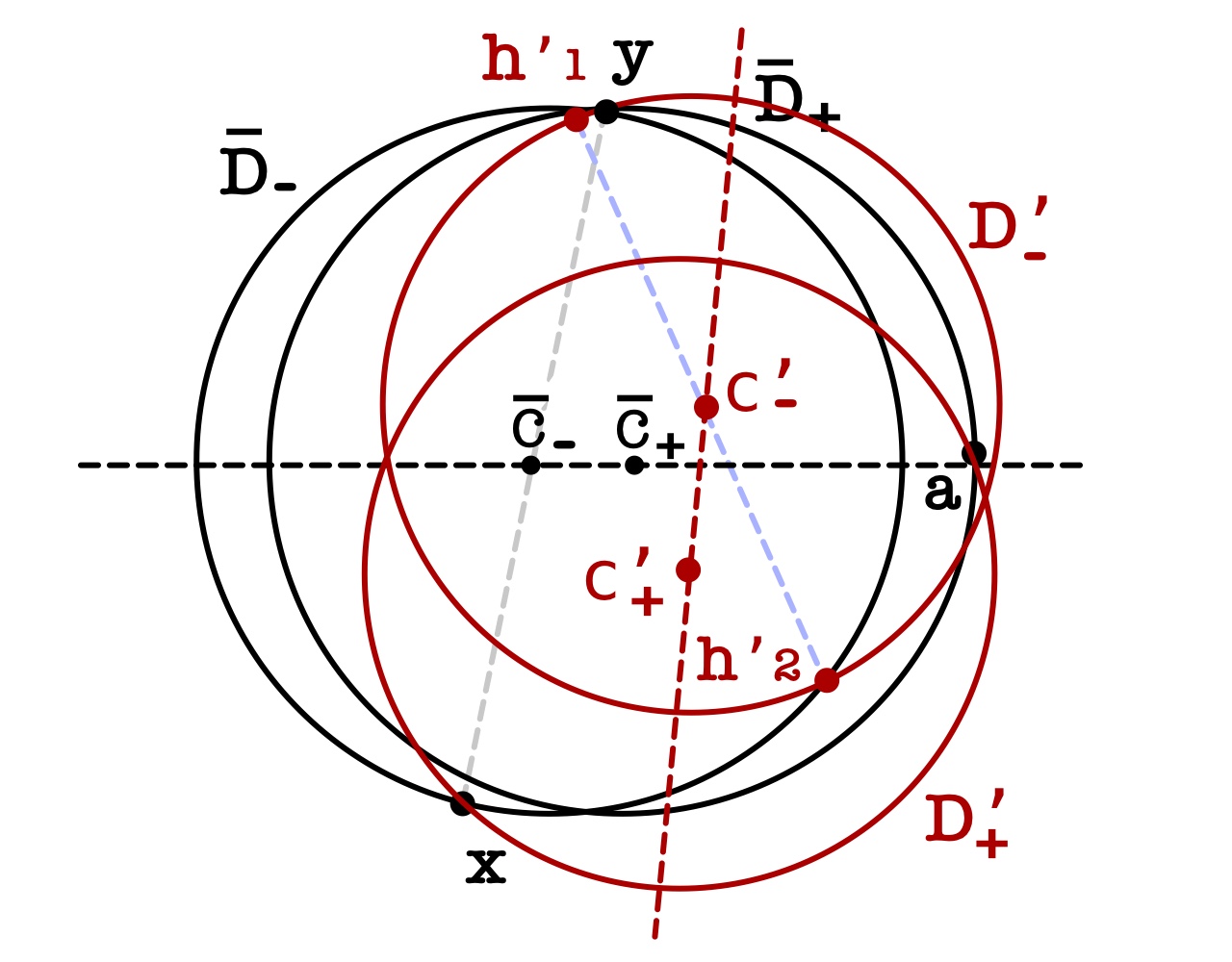}
\end{center}
\caption{Proof of Proposition \ref{prop_cpm_left}. The positions of $h'_1$ and $h'_2$ with respect to $c'_-$ (note that we relaxed the condition that $y$ should be covered by $D'_-$ to make the figure clear)}
\label{fig_prop91}
\end{figure}

In this sub-case, we assume that $x$ is below $line(\bar{c}_-,\bar{c}_+)$ and the case $x$ is above the line is similar.
\begin{prop}
$c'_+$ should be on the lower-right of $\bar{c}_+$. 
\label{prop_lower_right}
\end{prop}
\textbf{Proof.} $D'_+$ should cover $a,b$ and $x$. Based on Proposition \ref{prop_dom_dif_sides}, $a$ and $b$ should be on different sides of $line(\bar{c}_-,\bar{c}_+)$ and $a$ should be on the first quarter with respect to $\bar{c}_+$ and thus, outside of $\bar{D}_-$ (otherwise, because  of Proposition \ref{prop_dom_dif_sides} part 2, $b$ should be on the right side of $\bar{c}_+$. Now, $x\notin \bar{D}_+$ and $D'_+$ should contain the triangle $\bigtriangleup abx$ which contradicts $D'_+$ is smaller than $\bar{D}_+$). If $c'_+$ is on the lower-left of $\bar{c}_+$, $D'_+$ can't cover $a$ while it is smaller than $\bar{D}_+$. If $c'_+$ is on the top-left of $\bar{c}_+$, it should be above $line(a,\bar{c}_+)$ and $b$ should be below this line on $\partial \bar{D}_+$. This implies that $D'_+$ can't cover  $a$, $b$ and $x$ while being smaller than $\bar{D}_+$ (see Figure \ref{fig_prop8} for such situation). If $c'_+$ is on top-right $\bar{c}_+$, its distance from $x$ would be greater than the radius of $\bar{D}_+$ which is again not possible. $\hfill\square$
\\An immediate corollary of the above proposition is that $c'_-$ can't be on the lower-left of $c'_+$ otherwise, given the fact that $y$ is above $line(\bar{c}_-,\bar{c}_+)$ and $y\notin D'_+$ (otherwise $D'_+$ would have $a,b,x,y$ which are all dominating points of $\bar{D}_-$ and $\bar{D}_+$ and so can't be the smaller disk) $D'_-$ can't cover $y$.
\begin{obs}
$x$ and $\bar{c}_-$ should lie on different sides of $line(a,\bar{c}_+)$.
\label{obs_x_pos}
\end{obs}
The reason of the above observation is that because $a\notin \bar{D}_-$ and $x$ is below $line(\bar{c}_+,\bar{c}_-)$, if $x$ lies on the right side of  $line(a,\bar{c}_+)$, $axb$ would be a triangle for $\bar{c}_+$ which contradicts the assumption that $D'_+$ is smaller than $\bar{D}_+$. Also, $xa$ can't intersect $half\text{-}line (\bar{c}_+,\bar{c}_-)$ otherwise, $abx$ would be a triangle for $\bar{c}_+$ and again contradicts that $\bar{D}_+$ is the smaller disk. 
Henceforth, we denote the upper and lower intersection points of $\partial \bar{D}_-$ and $\partial \bar{D}_+$ by $I_1$ and $I_2$ respectively. Also, let $o$ be the mid-point of $\bar{c}_-\bar{c}_+$ and $I_3$ be the intersection of $half\text{-}line(\bar{c}_-,\bar{c}_+)$ and $\partial \bar{D}_+$.
\begin{prop}
$c'_-$ can't be on the right side of $c'_+$.
\label{prop_cpm_left}
\end{prop}
\textbf{Proof.} Suppose not and $c'_-$ is on the right side of $c'_+$. First note that $c'_-$ can't be on the lower-right of $c'_+$ otherwise $D'_+\cup D'_-$ can't cover both $y$ and $x$ while they are smaller (because of Proposition \ref{prop_dom_dif_sides} part 2 between $x$ and $y$). So, suppose that $c'_-$ lies on the top-right of $c'_+$. First note that $h'_2$ can't be above $line(\bar{c}_+,\bar{c}_-)$. To see why, let $\epsilon$ be the difference of the $x$-coordinates of $c'_+$ and $c'_-$. Also let $h$ be minimum difference of the $y$-coordinates of $c'_-$ and $c'_+$ in order to have $h'_2$ above $line(\bar{c}_+,\bar{c}_-)$. In order to keep the PCC, we need to have $h<\sqrt{\delta^2-\epsilon^2}$ but $h\geq \sqrt{r^2-(r-\delta-\epsilon)^2}$ (just assume that $c'_+$ lies on $line(\bar{c}_+,\bar{c}_-)$ and use the fact that $D'_+$ should cover $I_3$ to get the bound) which is not possible. On the other hand, $h'_1$ should come after $y$ in the counter-clockwise order because $D'_-$ needs to cover $y$. So, by adding $h'_1$ to the positive side, any disk covering $x,a,h'_1$ should also cover $h'_2$ which means its radius would be bigger than $M^+[\hat{i},\bar{j}]$ and so we can discard based on Theorem \ref{main_theorem} (see Figure \ref{fig_prop91}). $\hfill\square$ \vspace{-.5cm}
We know that $y\notin D'_+$. This is because if $y\in D'_+$, $D'_+$ would have $a,b,x,y$ which are the all dominating points of the solution of $M^+[\bar{i},\bar{j}]$ and so can't be the smaller disk. On the other hand, both $h'_1$ and $h'_2$ should be in $\bar{D}_-$. Let $h'_1$ be the first dominating point after $y$ in the counter-clockwise order. We recall that $h'_1$ should be outside $D'_+$ (otherwise, it contradicts Proposition \ref{prop_dom_dif_sides}). From Proposition \ref{prop_cpm_left}, we know that $c'_-$ is on the left side of $c'_+$. Also, note that $c'_-$ can't be below $c'_+$ otherwise $D'_-$ can't cover $y$ (consider Proposition \ref{prop_dom_dif_sides} between $x$ and $y$). So, the intersection point of $half\text{-}line(c'_-,c'_+)$ and $\partial D'_+$ should lie on the forth quarter with respect to $o$. This implies that one dominating point of $D'_+$ namely $h_1$ lies after $a$ in the angular counter-clockwise order and the other $h_2$ before $x$. \par
\begin{figure}[h]
\begin{center}
\includegraphics[scale=0.1]{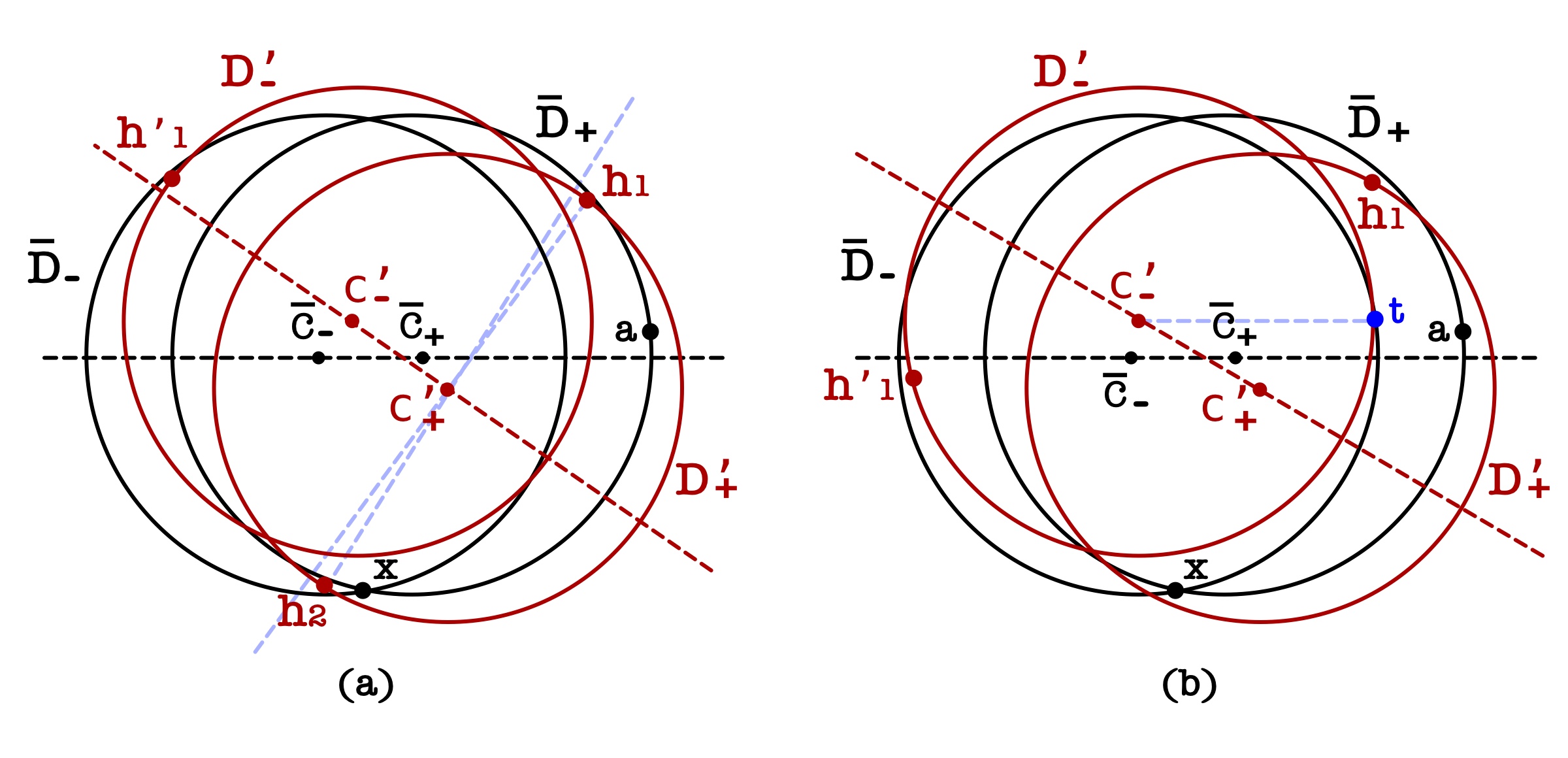}
\end{center}
\caption{(a) $h'_1$ is on the right side of $line(c'_+,c'_-)$. (b) when $t$ is inside $\bar{D}_-$, $dist(c'_-,c'_+)$ would be greater than $\delta$.}
\label{fig_leftright}
\end{figure}

We consider two cases. First, assume that $h'_1$ is on the right side of $line(c'_+,c'_-)$. In this configuration, because $c'_+$ is on the lower-right of $\bar{c}_+$ and Proposition \ref{prop_dom_dif_sides} for $x$ and $y$, after adding $h'_1$ to the positive side, we always have $\bigtriangleup h_1h_2h'_1$ around $c'_+$ and so we can discard (see Figure \ref{fig_leftright} (a)).
Now, assume that $h'_1$ is on the left side of $line(c'_+,c'_-)$. In this case, because $a$ is above $\bar{c}_+$ and $h_1$ is after $a$ in the order, the intersection of $half\text{-}line(h_1,c'_+)$ and $D'_-$ should be below $c'_+$. Now, if $h'_1$ is above $c'_+$, we again have triangle $\triangle h'_1h_1h_2$ for $c'_+$ and thus we can discard. Otherwise, $half\text{-}line(h'_1,c'_-)$ should intersect $\partial D'_-$ inside $\bar{D}_-$ (otherwise there would be no place for $h'_2$). Now, let $t$ be the point on $\partial D'_-$ with same $y$-coordinate as $c'_-$ on the left side of it (see Figure \ref{fig_leftright} (b)). Because $h'_1$ is below $c'_-$, $t$ should lie inside $D'_-$ but in order to have this condition $dist(c'_+,c'_-)$ should be greater than $\delta$ which is not possible.

\subsubsection{\underline{Sub-case 2: $\mathbf{x\in \bar{D}_+}$.}\\\\} 
\begin{figure}[H]
\begin{center}
\includegraphics[scale=0.1]{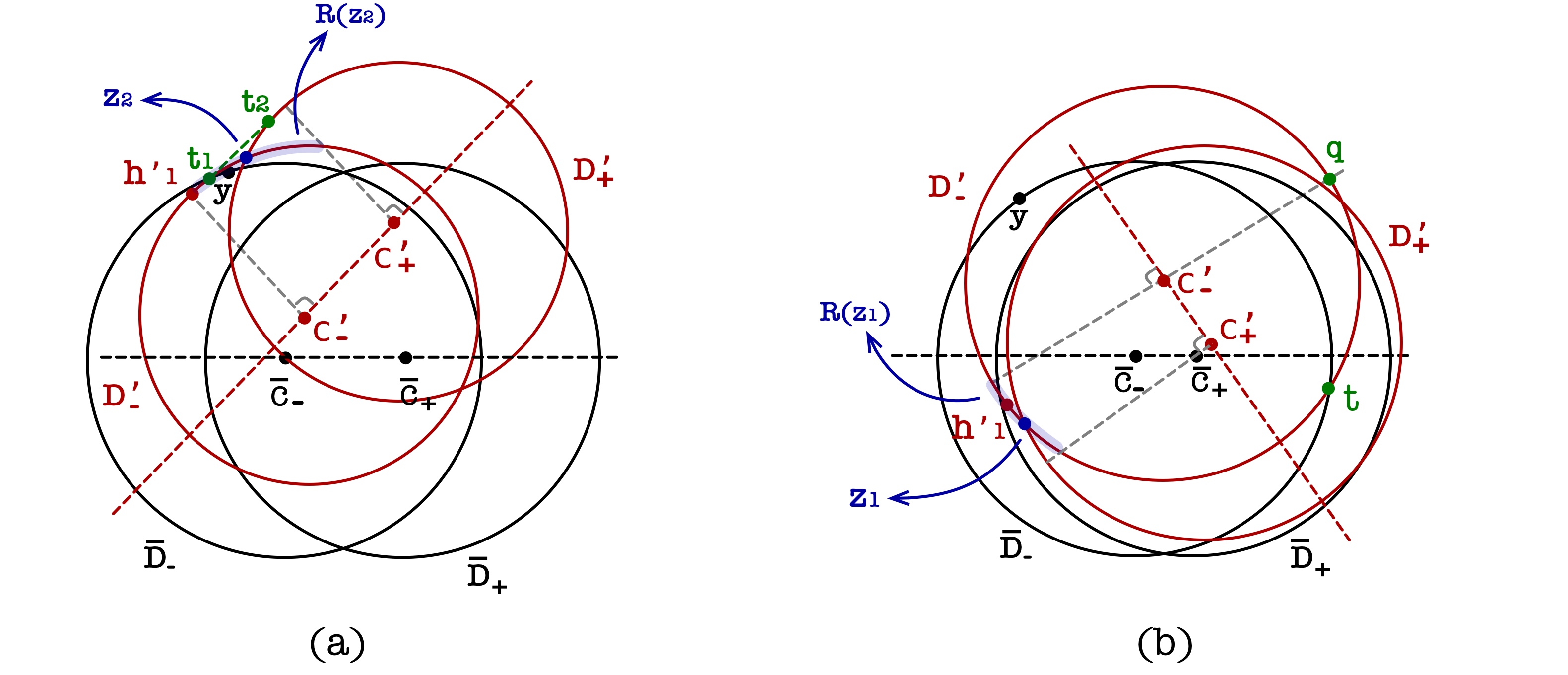}
\end{center}
\caption{An example of configuration of points for Proposition \ref{prop_regions}. Note that in this figure, we relaxed the condition that $x$ should be covered by $D'_+$ in order to illustrate situations where $h'_1$ lies inside $R(z_1)$ and $R(z_2)$.}
\label{fig_prop_regions}
\end{figure}

%Without lose of generality, we can assume that $x\in D^{\bar{i},\bar{j}}_-$ and the other case is similar.
In this sub-case, because $x\in \bar{D}_+$, $y$ should be outside of $\bar{D}_+$ and left side of $\bar{c}_-$ and indeed on $\partial convex\text{-}hull(P)$. Similar to the previous sub-section, let $h'_1$ be the first dominating point of $D'_-$ that appear after $y$ in the counter-clockwise order which should be outside $D'_+$. 
\begin{obs}
$c'_+$ lies on the right side of $\bar{c}_+$.
\label{obs_cp_right}
\end{obs}
This is because if $c'_+$ is on the left side of $\bar{c}_+$, there would be no place for $a$ and $b$ such that $ab$ does not intersect the $half\text{-}line(\bar{c}_+,\bar{c}_-)$ while keeping $D'_+$ the smaller disk.
%\begin{obs}
%The right intersection of the horizontal line passing from $c'_+$ and $D'_+$ is outside of $\bar{D}_+$.
%\label{obs_cpp_line_out}
%\end{obs}
%This is because $D'_+$ should cover both $a$ and $b$ and they are on different sides of $line(\bar{c}_+,\bar{c}_-)$ according to Proposition \ref{prop_dom_dif_sides} part 1. The above observation implies that if $h'_1\in D'_+$, then $h'_2$ should also lie below $c'_+$ and thus inside $D'_+$. So, $D'_+$ would have both $h'_1$ and $h'_2$ which is not possible which means that $h'_1$ is outside $D'_+$.
Similar to Proposition \ref{prop_cpm_left} we can assume that $c'_-$ is on the left side of $c'_+$ otherwise $h'_2$ should be covered by any disk covering $a,b$ and $x$. Now, let $z_1$ and $z_2$ be the two intersection points of $\partial D'_-$ and $\partial D'_+$ where $z_1$ appears first in the counter-clockwise order from $y$. Also, let $R(z_1)$ and $R(z_2)$ be the portions of $\partial D'_-$ between two perpendicular lines from $c'_-$ and $c'_+$ on $line(c'_-,c'_+)$ around $z_1$ and $z_2$ respectively (see Figure \ref{fig_prop_regions})
\begin{prop}
$h'_1$ does not intersect $R(z_1)$ and $R(z_2)$.
\label{prop_regions}
\end{prop}
\textbf{Proof.} We first show that $h'_1$ does not intersect $R(z_2)$. We proceed by contradiction and suppose that $h'_1\in R(z_2)$. Let $t_1$ be the intersection point of $\partial \bar{D}_-$ and $\bar{D}'_-$ which comes first after $y$. In this situation, $t_1$ should also be in $R(z_2)$. Also, let $t_2$ be the first intersection point of the half-line passing from $t_1$ parallel to $line(c'_-,c'_+)$ and $\partial D'_+$ (see Fig \ref{fig_prop_regions} (a)). Now, $t_2$ should be outside $\bar{D}_-$ because $t_2\in R(z_1)$ and the positive slope of $line(c'_-,c'_+)$ (in order to have have $t_1$ inside $R(z_2)$). On the other hand, $D'_+$ has both $t_2$ and $x$. If $t_2x$ intersect the $half\text{-}line(\bar{c}_-,\bar{c}_+)$, $|t_2x|$ should be greater than $M^+[\bar{i},\bar{j}]$ which is contradiction. Otherwise, $D'_+$ should have $a,b,x,t_2$ which again make it bigger than $M^+[\bar{i},\bar{j}]$ (consider a pair of disks with centers $\bar{c}_+$ and $\bar{c}_-$ and dominating points $\{a,b\}$ and $\{x,t_2\}$ respectively) .\par
\begin{figure}[h]
\begin{center}
\includegraphics[scale=0.12]{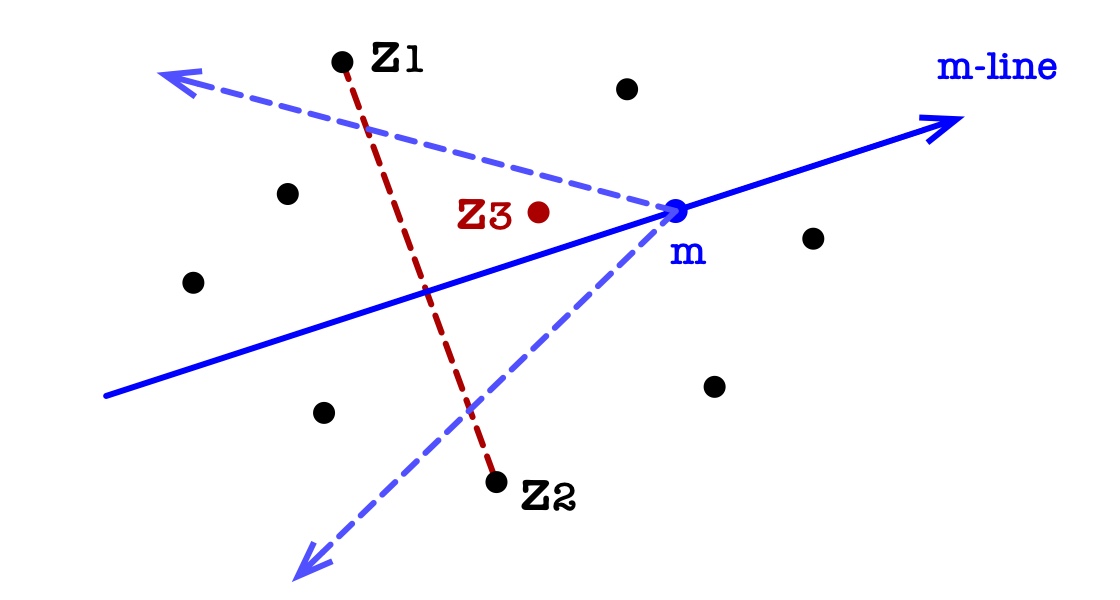}
\end{center}
\caption{A negative cone. $z_1,z_2$ make a cut for $z_3$.}
\label{fig_cone}
\end{figure}
\begin{figure}[h]
\begin{center}
\includegraphics[scale=0.12]{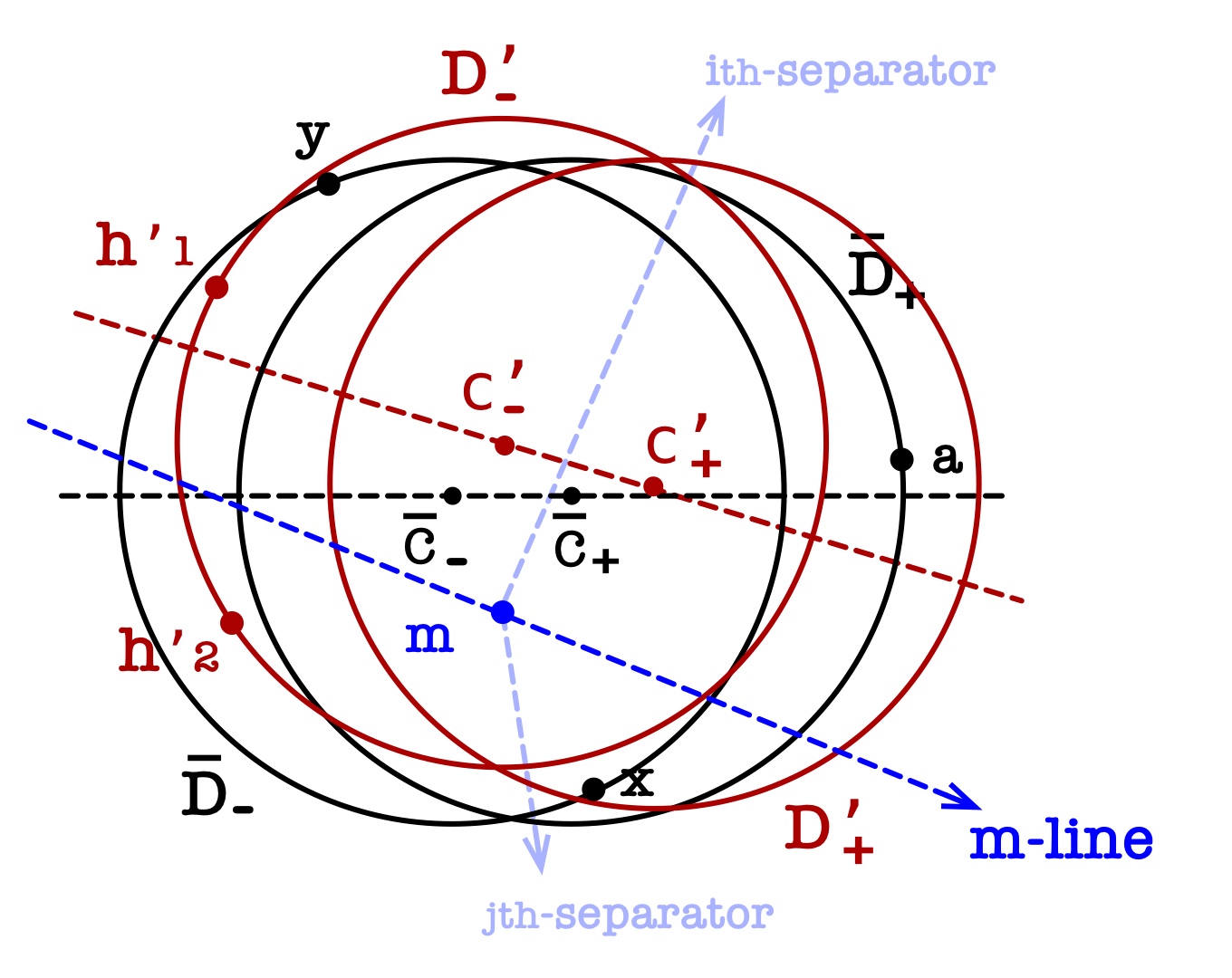}
\end{center}
\caption{In order to add $h'_1$ to the positive side, we first need to add $y$ to the positive side.}
\label{fig_last1}
\end{figure}
\begin{figure}[H]
\begin{center}
\includegraphics[scale=0.12]{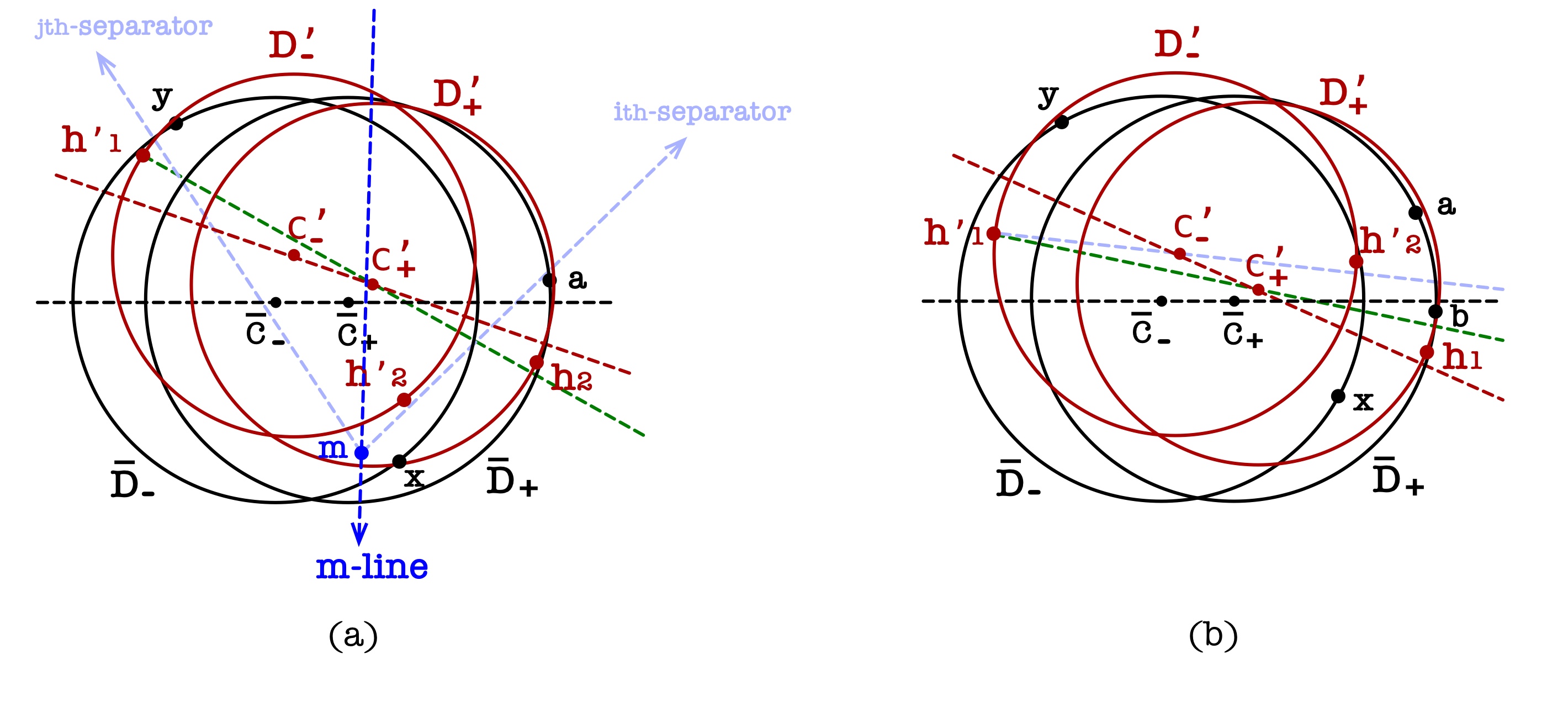}
\end{center}
\caption{$h'_1$ is on the right side of the $m$-line. (a) $h'_1$ is on the right side of $line(c'_+,c'_-)$. (b) $h'_1$ is on the left side of $line(c'_+,c'_-)$.}
\label{fig_right_right}
\end{figure}

Now, we prove that $h'_1$ can't intersect $R(z_1)$. Let $q$ be the last point of $R(z_2)$ in the counter-clockwise order. (see Figure \ref{fig_prop_regions} (b)). 
If $h'_1$ lies on $R(z_1)$, $h'_2$ needs to lie between $q$ and $\bar{D}_-$ on $\partial D'_-$ in order to not intersect $half\text{-}line(c'_-,c'_+)$. But, $q$ is outside $\bar{D}_-$ because the right intersection point $t$ of $\bar{D}_-$ and $D'_-$ is below $c'_-$ (this is because the PCC. Precisely, if $\zeta$ is the difference between the radii of $D'_-$ and $\bar{D}_-$, $c'_+$ should lie at least $\zeta$ to the right of $\bar{c}_+$ to cover $a$ and $b$. This implies that $c'_-$ should also lie at least $\zeta$ to the right of $c'_-$ to keep the PCC which make $t$ below $c'_-$) there is no place for $h'_2$ inside $\bar{D}_-$ which is contradiction.$\hfill\square$\\\\
Consider an $(i,j)$-partition. Let's call the (convex)cone obtained by $m$ as its vertex and the separator half-lines from $m$ as its sides the \textit{$(i,j)$-cone}. If the positive direction of the $m$-line is in the cone, we say the cone is positive otherwise we say it is negative. We say two points $z_1,z_2$ in $P^{i,j}_-$(resp. $P^{i,j}_+$) make a \textit{cut} for $z_3\in P^{i,j}_+$(resp. $z_3\in P^{i,j}_-$) in a positive(resp. negative) $(i,j)$-cone, if $z_1z_2$ intersects both the sides of the cone and does not separate $z_3$ from $m$ in the cone. See Figure \ref{fig_cone} for an example of a cone and a cut for it.
\begin{obs}
If two point $z_1$ and $z_2$ in $P^{i,j}_+$(resp. $P^{i,j}_-$) make a cut for a point $z_3$ in $P^{i,j}_-$(resp $P^{i,j}_+$) in a negative(resp. positive) $(i,j)$-cone, then if $z_3$ is not covered by $convex\text{-}hull(P^{i,j}_+)$ (resp. $convex\text{-}hull(P^{i,j}_-)$), then $m$ can not be covered by $convex\text{-}hull(P)$.
\label{obs_cut_convexhull}
\end{obs}
The reason of the above observation is that if $z_3$ is not covered by $convex\text{-}hull(P^{i,j}_-)$, there is a line that separates this convex hull and $m$. Now, by adding the points inside the cone to the convex hull, we just move this separating line closer to $m$ but this line can never reach $m$.\\\\ 
In order to discard a sub-row or a sub-column of $M^+[\hat{i},\bar{j}]$ according to Theorem~\ref{main_theorem}, we need to consider different configurations of the points in the $(\hat{i},\bar{j})$-partition. Because $x$ can be above or below $line(\bar{c}_-,\bar{c}_+)$, in order to cover these cases, we can assume that $x$ is below $line(\bar{c}_-,\bar{c}_+)$ but the $m$-line can take the both possible directions. %Note that if $a$ and $b$ are both $p$-type, we could discard the rest of the row of $M^+[\bar{i},\bar{j}]$ in the initial search and so we would not have marked $M^+[\hat{i},\bar{j}]$ so we assume that they have different types.\par %Similarly, if both $h_1$ and $h_2$ are $q$-type, we can discard all elements above $M^+[\hat{i},\bar{j}]$ and the theorem follows. So, henceforth, we assume these cases will not occur.\par
Let $h_1$ be the dominating point of $D'_+$ on the right side of $line(c'_+,c'_-)$ and $h_2$ be the other one. We proceed the following cases based on the position of the $m$-line with respect to $y$ and $h'_1$:\\\\
\textbf{1) Both $\mathbf{h'_1}$ and $\mathbf{y}$ are on the left side of the $\mathbf{m}$-line:} Based on Proposition \ref{prop_regions} $h'_1$ is on the boundary of $convex\text{-}hull(D'_-\cup D'_+)$ and because $x\in \bar{D}_+$, $y$ is also on the boundary of $convex\text{-}hull(\bar{D}_-\cup\bar{D}_+)$. Now, because all points in $P$ are covered by the two convex hulls, both $y$ and $h'_1$ should be on the boundary of $convex\text{-}hull(P)$. So, if $y,h'_1$ are on the left side of the $m$-line, because $m\in convex\text{-}hull(P)$ and $h'_1$ and $h'_2$ has different types (and so the $m$-line should pass between $h'_1$ and $h'_2$), when we add $h'_1$ to the positive side, we first need to add $y$ to the positive side and then $h'_1$ (see Figure \ref{fig_last1}). Which means that after adding $h'_1$, the positive side has $a,b,x,y$ which are the all dominating points of $\bar{D}_-$ and $\bar{D}_+$. This implies that any covering disk of them should have a radius greater than $M^+[\bar{i},\bar{j}]$ and so, we can discard the rest of the row or column of $M^+[\hat{i},\bar{j}]$ based on the type of $h'_1$.
\\\\ \noindent\textbf{2) $\mathbf{h'_1}$ and $\mathbf{y}$ are on the left and right sides of the $\mathbf{m}$-line respectively:}
Based on Proposition \ref{prop_regions} both $h'_1$ and $y$ are on the convex hull of the points. Now, one of $a$ or $b$ should also be on $convex\text{-}hull(P)$ which means it is not possible to add both of them to the positive side before either adding $y$ or $h'_1$ to the positive side which is not possible.\\\\
\textbf{3) $\mathbf{h'_1}$ is on the right side of the $\mathbf{m}$-line:}
We consider two cases: First, suppose that $h'_1$ lies on the right side of $line(c'_+,c'_-)$. Now, if $h_2$ lies on the right side of $line(h'_1,c'_+)$, then adding $h'_1$ makes a triangle $\triangle h'_1h_1h_2$ and we are done. Otherwise, $h_2$ should be on $\partial convex\text{-}hull(D'_-\cup D'_+)$ (because of Proposition \ref{prop_regions}) and so the points $h'_1$, $y$ and $h_2$ would be on $\partial convex\text{-}hull(P)$. Now, if $h_2$ and $h'_2$ has different types, $h_2$ and $h'_1$ should have a same type (because $h'_1$ and $h'_2$ had different types based on our assumption) and so on a same side of the $m$-line. In this case, because $y$ lies on the right side of $line(h_2,h'_1)$ and both $h'_1$ and $h_2$ lie on the right side of the $m$-line, because $h'_2$ should lie on the left side of the $m$-line, $y$ should also lie on the right side of the $m$-line. In this situation in order to add $h_2$ to the positive side, we need to add $h'_1$ and $y$ to the positive side first which is not possible. This argument implies that $h'_2$ and $h_2$ have a same type. Because $h_2$ needs to be added to the positive side before $h'_2$, $m$ should be on the left side of $line(h_2,h'_2)$ but in this situation, after adding $h'_1$ to the positive side, we would have negative cone $h'_1mh_2$ and $h'_1h_2$ makes a cut for $h'_2$ ($h'_2$ lies on the right side of $line(h'_1,c'_-)$ to satisfy Proposition \ref{prop_dom_dif_sides}) which implies that any disk covering $h'_1$ and $h_2$ should also cover $h'_2$ which means we can discard (see Figure \ref{fig_right_right} (a)). 
%then $h'_2$ should be on the right side of $line(h'_1,c'_-)$. Now, if $h_2$ is on the right side of this line, adding $h'_1$ will make triangle $h'_1h_1h_2$ and because $h'_1$ is outside $D'_-$ we can discard. Otherwise, $h_2$ and $h'_2$ should have a same type (if they don't have a same type, the $m$-line should pass between them and because both $h'_1$ and $h_2$ are on $convex\text{-}hull(D'_+\cup D'_-)$, $h_2$ would be added to the positive side before $h'_1$ which is impossible) and $h'_1mh_2$ makes a cone for $h'_2$ which means after adding $h'_1$, $h'_2$ is also need to covered by the positive disk and so again we can discard. \par
Now, suppose that $h'_1$ is on the left side of $line(c'_+,c'_-)$. %First note that $m$ should be on the left side of $line(a,x)$ because we need to first add $a$ to the positive side and then $x$ (in $(\bar{D}_-,\bar{D}_+)$ $a$ has been added to the positive side while $x$ is still in the negative side).  
Again, if $h_1$ is on the left side of $line(h'_1,c'_+)$, adding $h'_1$ makes triangle $\triangle h'_1h_1h_2$ and we can discard. So, we assume that $h_1$ is on the right side of $line(h'_1,c'_+)$. Note that here $h'_2$ can't lie above $c'_-$ because of PCC (similar to the argument of point $t$ in the proof of Proposition \ref{prop_regions}) which implies $c'_-$ is above $c'_+$ (otherwise, there would be no place for $h'_2$). Now, if $a\in\bar{D}_-$, by adding $h'_1$ to the positive side, any disk smaller than $\bar{D}_+$ covering $h_1,a,h'_1$ should also cover $h'_2$. This is because $h'_2$ is on the left side of $line(h'_1,c'_-)$ and the portion of the disk in the first quarter of $o$ would be outside of $\bar{D}_+$ (in order to cover $a$) and so we can discard. Let's assume that $a\notin \bar{D}_-$ and so on $\partial convex\text{-}hull(P)$. On the other hand, because $h_1$ is inside $\bar{D}_+$, $b$ should also lie on $convex\text{-}hull(P)$ (see Figure \ref{fig_right_right} (b)). Now, if $b$ is on the left side of $line(y,\bar{c}_-)$, $x,y$ make triangle for $\bar{c}_-$ with both $a,b$ (because $x\in \bar{D}_+$) and so, we could discard in the initial search. But if $b$ lies on the right side of $line(y,\bar{c}_-)$, because $h'_1$ is after $y$ in the counter-clockwise order, it is not possible to cover three points $h'_1,a,b$ with a radius smaller than $M^+[\hat{i},\bar{j}]$ and so again we can discard.

\end{document}